\documentclass[amsmath,amssymb,aps, pre,nofootinbib]{revtex4}
\usepackage{graphicx}
\usepackage{amsmath}
\usepackage{amsfonts}
\usepackage{amssymb}
\usepackage{fancyhdr}
\usepackage{xcolor}
\begin{document}
\title{Pinned, locked, pushed, and pulled traveling waves in structured environments}
\author{Ching-Hao Wang}
\thanks{equal contribution}
\affiliation{Department of Physics, Boston University, Boston, MA 02215}

\author{Sakib Matin}
\thanks{equal contribution}
\affiliation{Department of Physics, Boston University, Boston, MA 02215}

\author{Ashish B. George}
\thanks{equal contribution}
\email[corresponding author: ]{ashish.b.george@gmail.com}

\affiliation{Department of Physics, Boston University, Boston, MA 02215}

\author{Kirill S. Korolev}
\email[corresponding author: ]{korolev@bu.edu}
\affiliation{Department of Physics, Boston University, Boston, MA 02215}
\affiliation{Graduate Program in Bioinformatics, Boston University, Boston, MA 02215}

\date{\today}

%*************************************************************************
% Abstract
%*************************************************************************
\begin{abstract}

Traveling fronts describe the transition between two alternative states in a great number of physical and biological systems. Examples include the spread of beneficial mutations, chemical reactions, and the invasions by foreign species. In homogeneous environments, the alternative states are separated by a smooth front moving at a constant velocity. This simple picture can break down in structured environments such as tissues, patchy landscapes, and microfluidic devices. Habitat fragmentation can pin the front at a particular location or lock invasion velocities into specific values. Locked velocities are not sensitive to moderate changes in dispersal or growth and are determined by the spatial and temporal periodicity of the environment. The synchronization with the environment results in discontinuous fronts that propagate as periodic pulses. We characterize the transition from continuous to locked invasions and show that it is controlled by positive density-dependence in dispersal or growth. We also demonstrate that velocity locking is robust to demographic and environmental fluctuations and examine stochastic dynamics and evolution in locked invasions.
\end{abstract}

%\begin{keyword}
%reaction-diffusion \sep discrete \sep Allee effect \sep range expansion \sep mode locking \sep coupled map lattice \sep  plateau
%\end{keyword}
\maketitle
%*************************************************************************
% Introduction
%*************************************************************************
\section{Introduction}

Propagation of waves, fronts, and pulses is a recurrent theme in natural sciences~\citep{murray:mathematical_biology}. In evolution, they describe the geographic spread of a beneficial allele or the increase of fitness over time~\citep{fisher:wave, kolmogorov:wave, tsimring:wave}. In ecology, they capture epidemics, invasions of foreign species, and range shifts due to a climate change~\citep{network:wave, pateman:climate_invasion, phillips:toads}. In cell biology, traveling excitations describe electrical pulses in neurons, calcium waves in various tissues, and the assembly of large cytoskeletal complexes~\citep{murray:mathematical_biology, ramaswamy:wave, actin_hem:wave, heart:waves, nelson:biological_physics, tsimring:calcium_pinning, ishihara:aster_growth}. The list of applications also includes crystal growth, combustion fronts, and even the spread of entanglement in quantum mechanics~\citep{pelce:curved_fronts, schachenmayer:entanglement, jurcevic:entanglement}.

\begin{figure}[!h]
\begin{center}
\includegraphics[width=\linewidth]{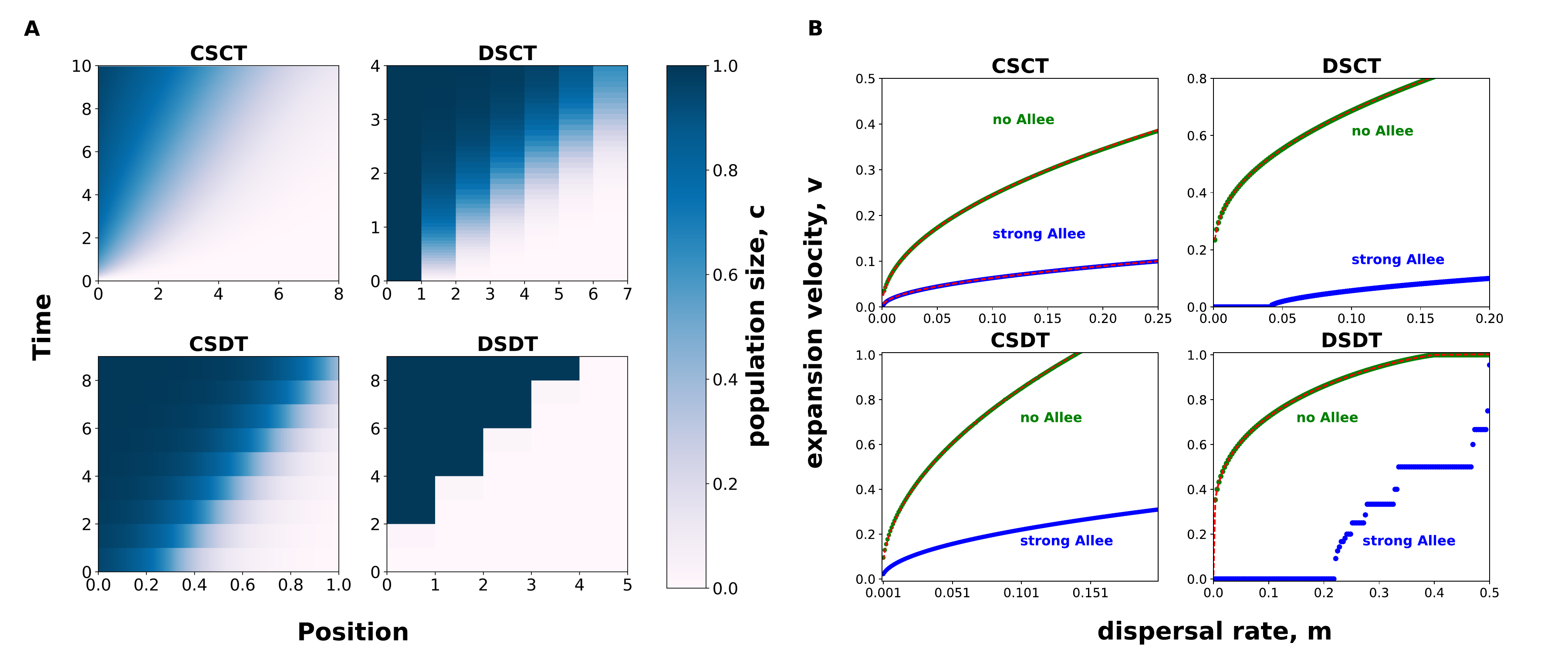}
\caption{\textbf{Discrete and continuous models predict distinct invasion dynamics.} Both panels compare four types of models that differ in whether space~(S) or time~(T) are treated as discrete~(D) or continuous~(C). \textbf{(A)} Species~(blue) expands into empty space~(white) starting from a step-like initial condition. Sharp changes in population density indicate discreteness in the underlying model. \textbf{(B)} The dependence of the expansion velocity~$v$ on the dispersal rate~$m$ for the same four models. When the space is continuous~$v(m)$ is a smooth function. The same behavior is observed in models with fragmented landscapes but continuous-time growth~(DSCT), except a critical dispersal rate might be necessary for the invasion to proceed. In contrast, the dependence of~$v$ on~$m$ could be highly non-analytic in models with discrete space and time~(DSDT), which manifests in constant velocities over a large range of dispersal rates. Note that all four models exhibit smooth~$v(m)$ in the absence of positive density-dependence~(Allee effect). Even for smooth~$v(m)$, there is a difference between the continuous-space models, for which~$v(m)\sim m^{1/2}$ for small~$m$, and discrete-space models, for which~$v(m)$ is more singular at the origin~\citep{fath:pinning}. Red dashed lines correspond to the analytical predictions described in the text. Parameter values are provided in Methods.}
\label{fig:table_schematic}
\end{center}  
\end{figure}

Given the diverse settings in which traveling fronts occur, numerous approaches have been developed to model front propagation in specific systems. Most of these approaches can be grouped into four classes depending on whether space and time are treated as discrete or continuous~(Fig.~\ref{fig:table_schematic}). Discrete time typically represents non-overlapping generations or periodic seasonal forcing, while discrete space accounts for habitat fragmentation.

Patchy habitats arise naturally due to a low density of locations suitable for growth, due to turbulent flows in marine environments~\citep{abraham:patchy_plankton, nelson:flow_prl_2012}, and due to internal ecological dynamics that create spatial patterns via Turing instabilities, ecological drift, and other mechanisms~\citep{wilson:coexistence, kefi:desert, korolev:mutualism, menon:diffusion}. Human development also leads to habitat fragmentation~\citep{fahrig:fragmentation} with row and grid planting of agricultural plants being extreme examples of regular spatial structures. In addition, habitats with regular arrangement of patches are often engineered using microfluidics or other technologies to create microcosm metapopulations for experimental study of collective behavior, ecology, and evolution using laboratory populations~\citep{datta:wave_splitting, gandhi:pulled_pushed, dai:nature, zhang:death_galaxy, active_matter:complex_environment, jorn:vortex_lattice}. Outside ecology, spatial structures arise due to crystal lattices, clustering of ion channels, or spatially periodic flow patterns in reaction chambers~\citep{tsimring:calcium_pinning, pelce:curved_fronts, jurcevic:entanglement, paoletti:epl_front_locking}. 

\subsection{Four classes of models with discrete or continuous space and time}
\textit{Continuous space, continuous time~(CSCT) models} arise naturally for chemical reactions and are typically formulated in terms of partial differential equations. In ecology and evolution, CSCT models describe organisms with overlapping generations living in a spatially homogeneous habitat. The simplest mathematical formulation of a CSCT model reads 

\begin{equation}
\frac{\partial c}{\partial t}=\frac{m}{2}\frac{\partial^2 c}{\partial x^2}+g(c),
\end{equation}

\noindent where $c(t,x)$ is the population density that depends position~$x$ and time~$t$; $m$ is the migration or dispersal rate; and~$g(c)$ is the growth rate of the population.

The only constraint on the functional form of~$g(c)$ is that there must be a stable equilibrium at~$c=K$, which is the carrying capacity. Mathematically, this constraint is expressed as~$g(K)=0$ and~$g'(K)<0$. An important property of~$g(c)$ is the limit~$\rho=\lim_{c\to0}g(c)/c$, which is the per capita growth rate at low density. When~$\rho<0$, a critical population size is required for the population to start growing, a phenomenon known as a strong Allee effect~\citep{allee:effect, courchamp:allee_review}. A weak Allee effect corresponds to positive~$\rho$ and~$g(c)/c$ increasing at low densities before declining to~$0$ at~$c=K$. When the per capita growth rate monotonically declines from~$\rho$ at~$c=0$ to zero at~$c=K$, there is no Allee effect. Allee effects arise due to a diverse set of mechanisms that include cooperative feeding, cooperative defense, mating, and difficulty in locating mates at low densities~\citep{allee:effect, courchamp:allee_review}.

Allee effects have profound implications for both the kinetics and evolutionary dynamics of range expansion~\citep{murray:mathematical_biology, fife:allee_wave, birzu:semipushed, korolev:arrest, korolev:wave_splitting, hastings:invasion_review, roques:allee_diversity}. Most of these effects can be understood from the classification of range expansions into either pulled or pushed~\citep{ birzu:semipushed, saarloos:review}. Pulled expansions~(or traveling waves) occur when there is no Allee effect or it is sufficiently weak. Then, the expansion dynamics are controlled by the rapid growth at the leading edge of the expansion. The velocity of pulled waves depends only on~$m$ and~$\rho$ and is given by the classic formula due to Fisher and Kolmogorov~$\sqrt{2m\rho}$~\citep{fisher:wave, kolmogorov:wave}. In contrast, the dynamics of pushed waves depends on the population growth throughout the expansion front, and, therefore, the expansion velocity depends on all the details of~$g(c)$. As a result, there is no general expression for the velocity of pushed waves. One exception to this is the following form of~$g(c)$ that has been extensively used in mathematical biology:

\begin{equation}
	\label{eq:cubic}
g(c)=g_0 c\left(1-\frac{c}{K}\right)\left(\frac{c}{K}-\frac{c_a}{K}\right),
\end{equation}

\noindent where $c_a$~controls the strength of an Allee effect, $g_0$~controls the growth rate, and~$K$ is the carrying capacity. When $c_a>0$, the Allee effect is strong, and~$c_a$ is the Allee threshold, i.e. the minimal density required for growth. The expansion velocity is known for all values of the Allee threshold~\citep{aronson:allee_wave, hadeler1975travelling, fife:allee_wave, korolev:wave_splitting} and is given by

\begin{equation}
v =\begin{cases} \sqrt{mg_0}\left(\frac{1}{2}-\frac{c_a}{K}\right), & c_a \ge -K/2\\
\sqrt{2mg_0|\frac{c_a}{K}|}, & c_a < -K/2.
\end{cases}
\label{eq:v_CSCT}
\end{equation}

\noindent Note that~$c_a=-K/2$ marks the transition between pulled and pushed expansions. For~$c_a>-K/2$, the shape of the front is also known exactly~\citep{aronson:allee_wave, hadeler1975travelling, fife:allee_wave, korolev:wave_splitting}. For both pulled and pushed expansions,~$v\sim\sqrt{m}$. We illustrate this dependence in Fig.~\ref{fig:table_schematic} using~$g(c)$ defined above. 

\textit{Continuous space, discrete time (CSDT) models} describe species with strong seasonality such as annual plants that spread in a spatially homogeneous landscape. These models are typically expressed in terms of integrodifference equations that specify how population densities change from generation to generation:

\begin{equation}
	\label{eq:ide}
	c_{t+1}(x)=\int_{-\infty}^\infty Q(x-x')f(c_t(x'))dx',
\end{equation}

\noindent where $c_t(x)$ is the population density as a function of position~$x$ at discrete generation~$t$; $Q(x-x')$ is the dispersal kernel that specifies the probability of dispersal to position~$x$ from position~$x'$; and $f(c_t(x))$ describes the growth dynamics. The kernels that are sufficiently short-range, e.g., exponential or Gaussian, result in an invasion front that propagates at a constant speed. Fat-tailed kernels,~e.g. with power law tails, could result in accelerating invasions~\citep{kot:csdt, wang:csdt, hallatschek:acceleration}.

To visualize the dynamics in CSDT models, we used exponential~(Laplace) dispersal kernel and a simple piecewise-linear growth model, which are defined below

\begin{equation}
	\label{eq:laplace}
Q(x)=\frac{1}{2\sqrt{m}} e^{- \frac{|x|}{\sqrt{m}}},
\end{equation} 

\begin{equation}
	f(c) = \left\{
	\begin{aligned}
		& rc, \;\; c<c^* \\
		& K, \;\; c\ge c^*,
	\end{aligned}
	\right.
	\label{eq:linear_model}
\end{equation}

\noindent where~$m$~quantifies dispersal distance,~$r$ is the growth rate at low densities, $K$~is the carrying capacity, and~$c^*$ is the density sufficient to reach the carrying capacity. When~$rc^*=K$, the growth function is continuous, and there is no Allee effect because the per capita growth rate~$f(c)/c$ monotonically decreases with population density. When~$rc^*<K$, there is a jump in~$f(c)$ at~$c=c^*$. This jump corresponds to an increase in the growth rate due intraspecific facilitation, which requires a certain population density~\citep{allee:effect, courchamp:allee_review}. Because of the jump,~$f(c)/c$ is non-monotonic and, therefore, exhibits an Allee effect. When~$r>1$, small populations always grow, and the Allee effect is said to be weak. Strong Allee effect occurs for~$r<1$. In this regime, small populations go extinct unless they are rescued by immigration from nearby patches. 

Similar to CSCT models, one can distinguish between pulled and pushed waves. Similar to CSCT models, the velocity of pulled waves depends only on the growth dynamics at low densities~$r=\lim_{c\to0}f(c)/c$. The role of dispersal is, however, more complex and~$v$ depends on all the details of~$Q(x)$. For the Laplace kernel defined above, the velocity of pulled waves is given by; see Appendix~A and Ref.~\citep{kot:csdt, wang:csdt}:

\begin{eqnarray}
v =  \min_{\kappa>0}\left\{  \frac{  \ln\left[ \frac{r}{1-m \kappa^2} \right] }{\kappa}\right\}.
\end{eqnarray}

\noindent Because one can always rescale the spatial coordinate to eliminate~$m$, the expansion velocity is proportional to $\sqrt{m}$ for both pulled and pushed waves. We illustrate this dependence in Fig.~\ref{fig:table_schematic} using~$Q(x)$ and~$f(c)$ defined above.

\textit{Discrete space, continuous time (DSCT) models} are appropriate for fragmented landscapes with little or no seasonal forcing. The dynamics are described by a set of ordinary differential equations coupled by dispersal, which, in the simplest case, occurs only between the nearest patches:

\begin{equation}
	\label{eq:dsct}
\frac{d c_x}{d t}=\frac{m}{2}(c_{x+1}-2c_x+c_{x-1})+g(c_x).
\end{equation} 

The dynamics of DSCT expansions is shown in Fig.~\ref{fig:table_schematic} using~$g(c)$ defined by Eq.~(\ref{eq:cubic}). Note that~$v$ is no longer proportional to~$\sqrt{m}$ in either pulled or pushed waves. The velocity of pulled waves depends only on~$m$ and~$\rho=\lim_{c\to0}g(c_x)/c_x$ and is derived in Appendix~A. In contrast to CSCT and CSDT models, the velocity of pushed waves in DSCT models can be pinned at~$v=0$ for a range of~$m$. Velocity pinning requires a strong Allee effect and occurs because low values of~$m$ are insufficient to push the populations at the range margin over the Allee threshold~\citep{keitt:pinning, fath:pinning}. 

\textit{Discrete space, discrete time (DSDT) models} form the remaining class that describes seasonal growth in fragmented landscapes. In this case, the dynamics are governed by difference equations, which are also known as coupled map lattices and cellular automata:

\begin{equation}
	c_{t+1,x} = f\left( \frac{m}{2}c_{t,x-1} + (1-m)c_{t,x} + \frac{m}{2}c_{t,x+1} \right).
	\label{eq:dsdt}
\end{equation}

\noindent In addition to natural populations, DSDT models capture the dynamics in some experimental studies~\citep{datta:wave_splitting, gandhi:pulled_pushed}. Moreover, DSDT model underlie many simulations because discretization is frequently deployed in numerical methods. 

The dynamics of DSDT models are illustrated in Fig.~\ref{fig:table_schematic} using~$f(c)$ defined by Eq.~(\ref{eq:linear_model}). The velocity of pulled waves is derived in Appendix A and depends only on~$r$ and~$m$. The velocity of pushed waves exhibit a striking phenomenon of velocity locking: There is a discrete set of velocities at which~$v$ remains constant despite variation in~$m$~\citep{carretero:locking_physica,carretero:locking_pre, fernandez:jsp, coutinho:extended, coutinho:convolution, coutinho:monotone, turzik:stability}. A detailed study of the ecological and evolutionary implications of the locking phenomenon is the central topic of this paper.

\subsection{Locking in DSDT models}
\noindent Despite significant differences, the models in all four classes can describe the propagation of a traveling front. As a result, the choice of the model is often dictated not only by its relevance to a specific population, but also by its mathematical or computational tractability~\citep{petrovskii:exactly_solvable, mistro:complexity_allee, lewis:chamomile}. Reaction-diffusion equations are commonly used to obtain analytical insights into population dynamics~\citep{murray:mathematical_biology, fife:allee_wave, birzu:semipushed, korolev:arrest, korolev:wave_splitting, hastings:invasion_review, roques:allee_diversity}, and integrodifference equations are preferred when it is necessary to account for long-range dispersal~\citep{kot:csdt, wang:csdt}. Models with discrete time or space also have certain advantages; for example, they provide a convenient way to introduce stochastic dynamics or use satellite data on natural landscapes~\citep{kimura:ssm, korolev:rmp, lewis:chamomile}. 

Some phenomena, however, arise most naturally only in a specific model class. One important example in invasion pinning~(Fig.~\ref{fig:table_schematic}), which occurs when the front gets stuck at a particular patch or landscape heterogeneity~\citep{keitt:pinning, fath:pinning}. Invasion pinning requires habitat fragmentation and the presence of a strong Allee effect, i.e. a threshold density below which the growth rate is negative. Under these conditions, a critical dispersal rate is required to initiate population growth in neighboring patches. Above the critical dispersal rate, invasion proceeds as in continuous-space models, but, below it, invasion front is pinned. In models where both time and space are discrete, invasions can become not only pinned, but also locked. Locked fronts advance periodically in a pulsed fashion, and their velocities assume only a discrete set of values, which are insensitive to small changes in dispersal and growth rates~(Fig.~\ref{fig:table_schematic}).

Although locking of traveling fronts may seem peculiar, it has been observed for a Belousov-Zhabotinsky reaction spreading in a periodic array of periodically-driven vortices~\citep{paoletti:epl_front_locking}. Moreover, the characteristic pattern of plateaus shown in~Fig.~\ref{fig:table_schematic} was found in a variety of physical systems including arrays of Josephson junctions~\citep{basler:staircase_josephson}, chemical reactions~\citep{maselko:staircase_chemical}, quantum gases~\citep{lan:staircase_quantum_gas}, and crystal structures of alloys~\citep{komura:staircase_material}. This pattern, dubbed a Devil's staircase~\citep{bak:devils_staircase}, is a manifestation of a phenomenon known as mode locking in nonlinear sciences. In the simplest case, mode locking occurs when the observed frequency of a nonlinear oscillator equals a rational number times the frequency of the external drive. The plateaus are regions of drive amplitude that correspond to the same ratio of the actual oscillator frequency and the frequency of the drive.

The theory of mode locking has been used in the context of coupled map lattices to explain the locking of invasion velocities, and some rigorous results are available on the existence and properties of locked fronts~\citep{carretero:locking_physica,carretero:locking_pre, fernandez:jsp, coutinho:extended, coutinho:convolution, coutinho:monotone, turzik:stability}. Previous studies, however, lacked ecological context and focused exclusively on bistable maps. Therefore, they did not explore the transition from continuous to locked fronts as the strength of an Allee effect or other ecological parameters are varied~(Fig.~\ref{fig:table_schematic}). Understanding this transition and the differences between DSDT and other models are the primary goals of this paper. We show that velocity locking requires positive density-dependence in growth or dispersal, but bistability is not necessary. In addition to locked and pinned waves, we report two types of continuous expansions in DSDT models. These types are analogous to pulled and pushed waves in reaction-diffusion equation~\citep{saarloos:review, birzu:semipushed, gandhi:pulled_pushed}. Our results further demonstrate that velocity locking occurs both in one and two spatial dimensions and is robust to significant levels of demographic and environmental fluctuations. The effects of velocity locking on stochastic front wandering and evolutionary dynamics are also discussed.

%**************************************************************************
% Results
%**************************************************************************
\section{Results}

\begin{figure}
\begin{center}
\includegraphics[width=\linewidth]{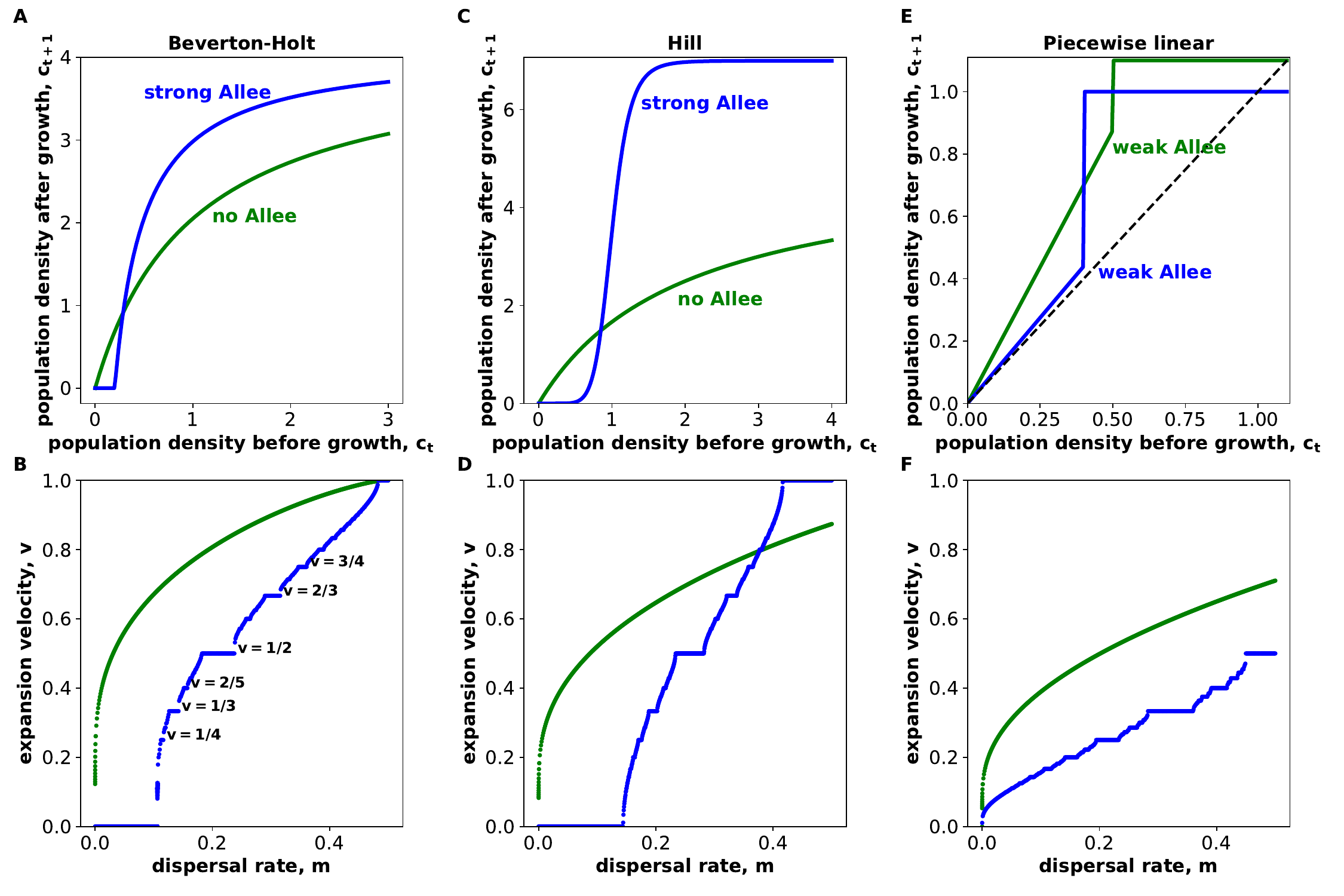}
	\caption{\textbf{Velocity locking occurs in many ecological models.} The growth functions~$f(c)$ for three ecological models are shown in the top three panels~\textbf{(A, C, E)}. The bottom panels~\textbf{(B, D, F)} show the dependence of the invasion velocity on the migration rate for the corresponding growth model. Panels~B and~D illustrate the fact that velocity locking occurs in the presence of a strong Allee effect, but not when an Allee effect is absent. Panels~E and~F show that velocity locking can occur for a weak Allee effect, but only if it is sufficiently large. For Beverton-Holt model, we used~$A=4.1$,~$B=0.3$, and~$c^*=0.2$ for the strong Allee effect conditions and~$A=4.1$,~$B=1$, and~$c^*=0$ for the conditions without an Allee effect. For the Hill model, we used~$A=7$,~$B=1$, and~$n=8$ for the strong Allee effect conditions and~$A=5$,~$B=2$, and~$n=1$ for the conditions without an Allee effect. For the piecewise-linear model, we used~$r=1.75$,~$K=1.1$, and~$c^*=0.5$ for the slight weak Allee effect conditions~(green curve) and~$r=1.1$,~$K=1.0$, and~$c^*=0.4$ for the substantial weak Allee effect conditions~(blue curve).}
	\label{fig:locking_common}
\end{center}  
\end{figure}

\subsection{Velocity locking occurs in many ecological models}
We studied a variety of ecological growth models to determine whether velocity locking is a generic phenomenon. Although the piecewise-linear model in Eq.~(\ref{eq:linear_model}) is a reasonable approximation of ecological dynamics, it not clear whether velocity plateaus in Fig.~\ref{fig:table_schematic} arise due to discontinuities in~$f(c)$. To answer this question, we considered two alternative growth functions that had complementary properties and covered a wide range of ecological scenarios. Note, however, that we only considered growth functions that describe a steady approach to the carrying capacity and do not lead to chaotic or periodic dynamics on their own.

The Beverton-Holt model with offset~\citep{beverton_holt:model, chen:allee_model} is commonly used to model population dynamics, e.g. of fisheries, and is specified by the following equation:

\begin{equation}
	f(c) = \left\{
	\begin{aligned}
		& \frac{A (c-c^*)}{B + (c-c^*)}, \;\; c>c^* \\
		& 0, \;\; c\le c^*,
	\end{aligned}
	\right.
	\label{eq:bh}
\end{equation}

\noindent where~$A$ sets the carrying capacity, $B$~sets the density at which intraspecific competition starts to significantly affect population growth, and~$c^*$ sets the magnitude of the Allee effect.  

Although~$f(c)$ in the Beverton-Holt model is continuous, it has a discontinuous derivative at~$c^*$. To exclude the possibility that this non-analyticity is responsible for velocity locking, we considered an infinitely differential map, known as Hill function in molecular biology~\citep{hill:function}. Hill-like functions have also been used to model ecological populations~\citep{yakubu:allee_model}. The simplest Hill function reads

\begin{equation}
	f(c) = \frac{A c^n}{B + c^n},
	\label{eq:hill}
\end{equation}

\noindent where~$A$ and~$B$ play the same role as in the Beverton-Holt model, and~$n$ controls the strength of an Allee effect. The Allee effect is absent for~$n=1$ and increases with~$n$ for~$n>1$. 

Both Beverton-Holt and Hill models exhibited velocity locking, which is evident from the plateaus of~$v(m)$ shown in~Fig.~\ref{fig:locking_common}. Thus, velocity locking is not caused by singularities in the growth function, although velocity plateaus are larger for discontinuous or rapidly varying~$f(c)$. The main conclusion that we draw from Fig.~\ref{fig:locking_common} is that velocity locking is robust to the choice of modeling assumptions, control parameters, and differences in species ecology. Therefore, it should be expected for a generic growth model.

\subsection{Locked invasions are periodic}
\begin{figure}[!h]
\includegraphics[width=\linewidth]{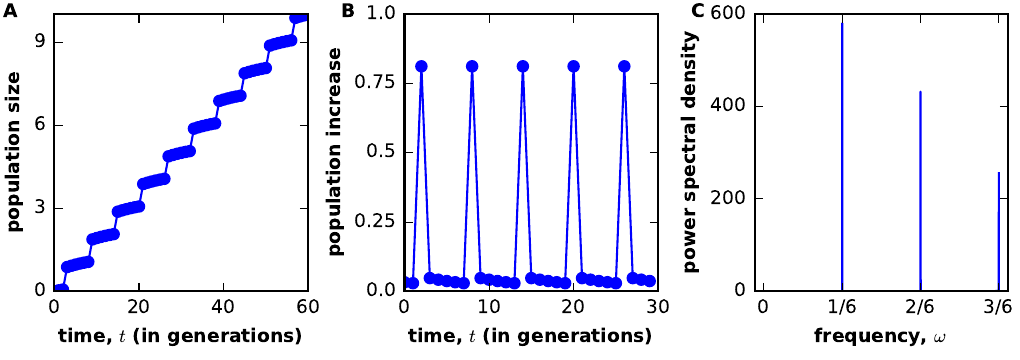}
\caption{\textbf{Locked fronts advance by periodic pulses.} \textbf{(A)} shows how the total population size of the invader increases with time for~$v=1/6$ velocity plateau in the piecewise-linear growth model. The staircase-like increase of the population size indicates that the invasion is pulsed. This is more clearly visible in \textbf{(B)}, where the change in the population size per generation is plotted. The pattern is clearly periodic, and most of the growth occurs only during one of the generations within a full cycle. Thus, the invader spreads in pulses that repeat every six generations. \textbf{(C)} confirms the periodicity of the invasion by showing the power spectrum of the population size change relative to its mean over the period. The peaks at the fundamental frequency of $1/6$ and higher harmonics are clearly visible. Here,~$m=0.110$,~$r=0.93$,~$K=1$, and~$c^*=0.22$.}
\label{fig:pulsed} 
\end{figure}

Locking of the velocity into a specific value results in a periodic and often pulsed propagation of the invasion front~(Fig.~\ref{fig:pulsed}). The pulsations occur because several generations of slow growth and dispersal are necessary to reach the density at which intraspecific facilitation enables rapid growth. After this rapid growth, the density profile assumes exactly the same shape as in the beginning of the cycle.

\begin{figure}[!h]
\begin{center}
\includegraphics[width=\linewidth]{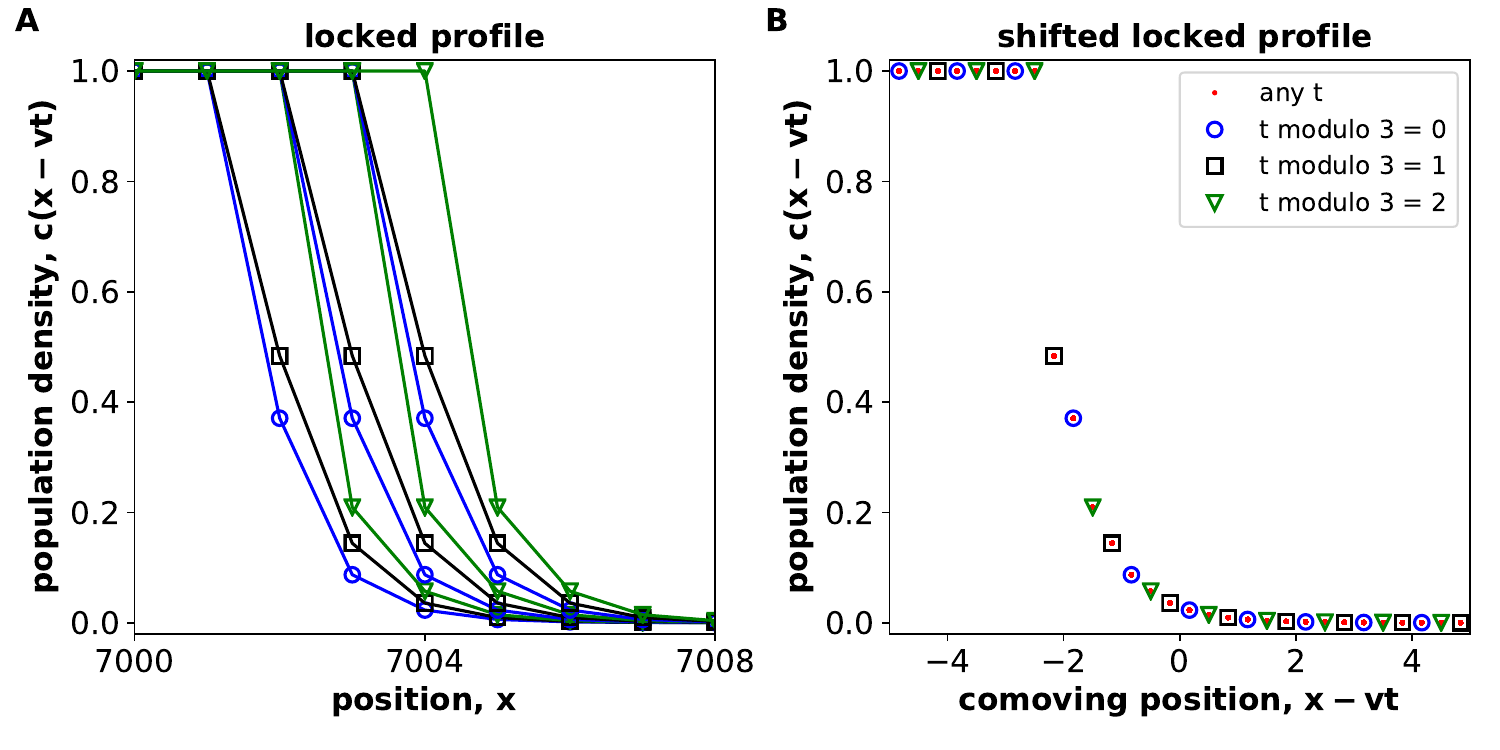}
	\caption{\textbf{Locked invasions have discrete profiles that repeat periodically.} \textbf{(A)} shows the population density profile at nine consecutive generations for~$v=1/3$ plateau in the piecewise-linear model. To highlight the fact that density profiles repeat periodically, we used different colors and symbols for different generations modulo~$3$. \textbf{(B)} Density profiles collected over~$10^3$ generations were shifted by~$-vt$ to transform them into the reference frame comoving with the invasion. The resulting profile is discrete with only a countable number of distinct values of~$c$. Here,~$r=1.1$,~$K=1$,~$c^*=0.5$, and~$m=0.4$. }
	\label{fig:understanding_locking}
\end{center}  
\end{figure}

In general, locked fronts advance by~$p$ patches every~$q$ generations~\citep{carretero:locking_physica,carretero:locking_pre}~(Fig.~\ref{fig:understanding_locking}), and there are only~$q$ distinct density profiles: one for each generation from~$t$ to~$t+q$. As a result, invasion velocities take only rational values,~$v=p/q$, when measured in units of inter-patch distances per generation~(Fig.~\ref{fig:locking_common}). The periodicity of front motion remains invariant within the velocity plateau even though the shape of the front does change with the model parameters. 

\begin{figure}[!h]
\begin{center}
\includegraphics[width=\linewidth]{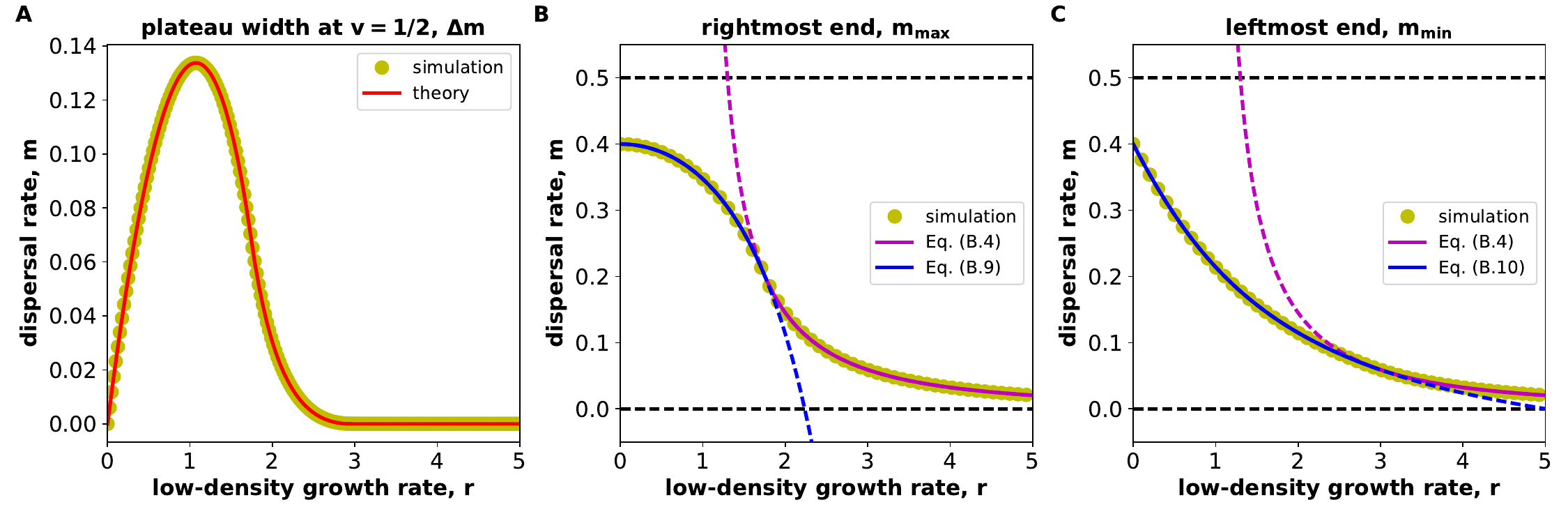}
	\caption{\textbf{Exact solution for plateau width confirms velocity locking.} \textbf{(A)} The width of the plateau at~$v=1/2$ in simulations (yellow dots) matches the theoretical prediction~(red line) that assumes that the front shape repeats periodically~(see Appendix B). This agreement confirms that~$v$ is exactly rather than approximately constant within the plateau. The theory predicts the rightmost~\textbf{(B)} and leftmost~\textbf{(C)} ends of the plateau correctly. Each end of the plateau is given by one of the two equations plotted in each panel. An additional condition determines which of the two equations is applicable~(see Appendix B). Solid lines indicate when the equation applies and dashed lines indicate when the equation does not apply. We used~$K=1$,~$c^*=0.2$, and varied~$m$ between~$0$ and~$0.5$. }
	\label{fig:plateau_width_prediction}
\end{center}  
\end{figure}

To verify the periodic nature of locked waves, we obtained an exact solution for the invasion dynamics inside~$v=1/2$ plateau for the piecewise-linear growth model~(see Appendix B). Our solution matches both the population density profiles and the locations of the velocity plateaus obtained in simulation~(Fig.~\ref{fig:plateau_width_prediction}). The fact that Eq.~(\ref{eq:dsdt}) admits a moving front with the same periodicity over a region of model parameters firmly establishes that~$v$ is exactly rather than approximately constant within the plateau. More importantly, this observation confirms that velocity locking is a result of the synchronization between the invasion dynamics and the spatio-temporal periodicity of the habitat.

\subsection{Pulled waves propagate without locking in DSDT models}
Not all invasions in DSDT models are locked; see Figs.~\ref{fig:table_schematic}~and~\ref{fig:locking_common}. In fact, it can be rigorously demonstrated that~$v(m)$ is a smooth function for pulled waves that are driven by the growth and dispersal at the very tip of the invasion front. Invasions are guaranteed to be pulled when there is no Allee effect, and, therefore, the per capita growth rate is maximal at the leading edge of the front~\citep{kolmogorov:wave, saarloos:review, lewis:chamomile, lui:dsdt_velocity}. Because~$c$ is small at the front, nonlinear terms in~$f(c)$ are negligible, and the invasion velocity can be computed exactly by linearizing Eq.~(\ref{eq:dsdt}) and then solving it via Laplace and Fourier transforms~\citep{saarloos:review}; see Appendix A. 

The main result of this calculation is that pulled waves in DSDT models behave as in CSCT models. Specifically, the population density at long times depends only on~$x-vt$ and is described by a continuous function~$c(x-vt)$ with~$v$ given by

\begin{equation}
	v = \min_{\kappa>0}\left\{\frac{\ln\left(\rho[1 + m (\cosh\kappa-1)]\right)}{\kappa}\right\},
	\label{eq:v_min}
\end{equation}

\noindent where~$\rho=\lim_{c\to0}f(c)/c$.

\begin{figure}[!h]
\begin{center}
\includegraphics[width=\linewidth]{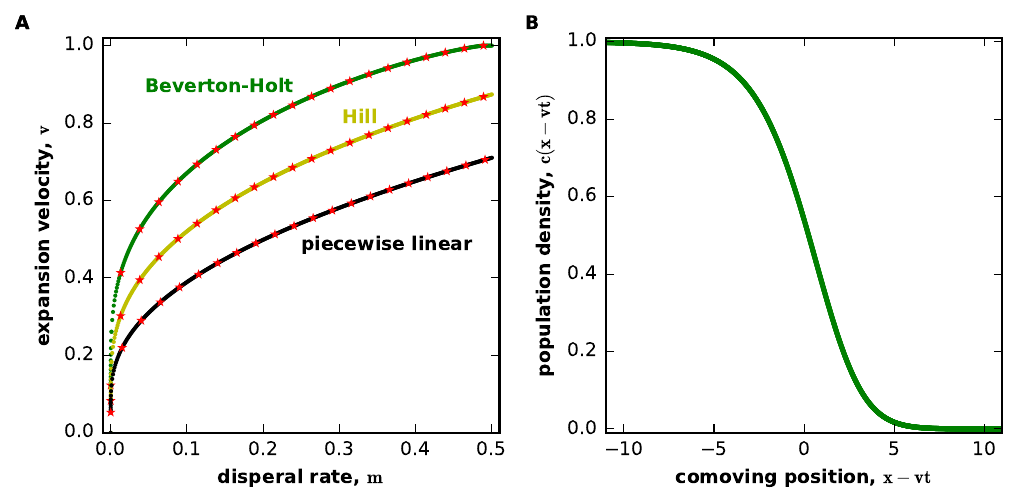}
	\caption{\textbf{Pulled waves propagate without velocity locking.} \textbf{(A)} The dependence of the invasion velocity on the dispersal rate is shown for Beverton-Holt, Hill, and the piecewise-linear growth models from Fig.~\ref{fig:locking_common}. The theoretical prediction for pulled waves from Eq.~(\ref{eq:v_min}) is shown with red stars. The excellent agreement between the theory and simulations confirms that in DSDT models species can spread without velocity locking. The parameters are the same as in Fig.~\ref{fig:locking_common}. \textbf{(B)} Pulled waves have a continuous density profile in the comoving reference frame. Population densities were obtained in the Beverton-Holt model for~$10^3$ generations and then shifted by~$-vt$ to transform them into the comoving reference frame. The observed collapse indicates that, in steady state,~$c(t,x)=c(x-vt)$ for pulled waves. In this panel, we used~$A=3$,~$B=2$,~$c^*=0$, and~$m=0.5$.}
	\label{fig:pulled_works}
\end{center}  
\end{figure}

In the absence of an Allee effect, numerical simulations show excellent agreement with these predictions. The invasion velocities predicted by Eq.~(\ref{eq:v_min}) perfectly matched the simulation results for the piecewise-linear, Beverton-Holt, and Hill model~(Fig.~\ref{fig:pulled_works}A). In addition, the observed population densities formed a continuous profile after a shift by~$vt$ to transform them into the reference frame comoving with the invasion~(Fig.~\ref{fig:pulled_works}B). Note that, for locked waves, the shift to the comoving reference frame produces a discrete profile~(Fig.~\ref{fig:understanding_locking}B) because there are only~$q$ distinct profiles. We therefore conclude that pulled waves are not locked, and DSDT models support both periodic and aperiodic invasions.

\subsection{Pushed waves: A second type of unlocked expansions in DSDT models}
\begin{figure}[!h]
\begin{center}
\includegraphics[width=0.5\linewidth]{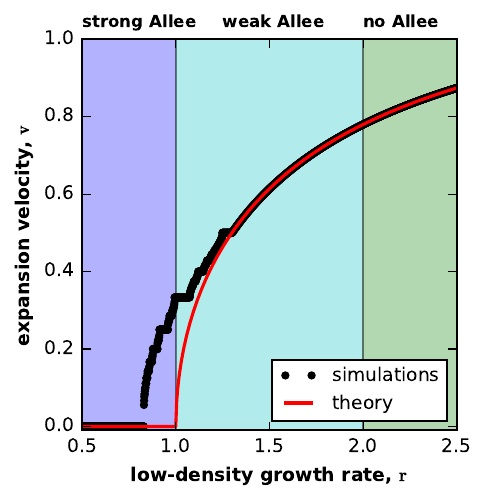}
\caption{\textbf{Expansions stop being pulled at intermediate strengths of a weak Allee effect.} Black dots show the dependence of the invasion velocity on the low density growth rate for the piecewise-linear model, and the red line shows the prediction of Eq.~(\ref{eq:v_min}).  For high~$r$, the data and theory overlap indicating that waves are pulled. For smaller~$r$, there is disagreement and clear signs of velocity locking. The transition between the two regimes occurs at~$r\approx 1.3$, which is substantially lower than the critical growth rate that marks the onset of a weak Allee effect~($r= 2.0 $). Here,~$K=1$,~$c^*=0.5$, and~$m=0.5$. }
	\label{fig:transition}
\end{center}  
\end{figure}

To understand the transition from locked to unlocked fronts, we examined how the invasion dynamics change with the strength of an Allee effect. For reaction-diffusion waves in CSCT models, a critical strength of the Allee effect is necessary to transform pulled waves into a new spreading regime~\citep{saarloos:review, birzu:semipushed, gandhi:pulled_pushed}. Invasions in this regime are known as pushed waves because nonlinearities in~$f(c)$ contribute to front propagation and, therefore, the spreading velocities are greater than predicted by Eq.~(\ref{eq:v_min}). We found similar behavior in DSDT models. Figure~\ref{fig:transition} shows that Eq.~(\ref{eq:v_min}) remains valid even in the presence of a substantial Allee effect, but the observed velocities become greater than the prediction of the linear theory once a critical strength of the Allee effect is reached.

\begin{figure}[!h]
%\begin{center}
\centerline{  \includegraphics[width=\linewidth]{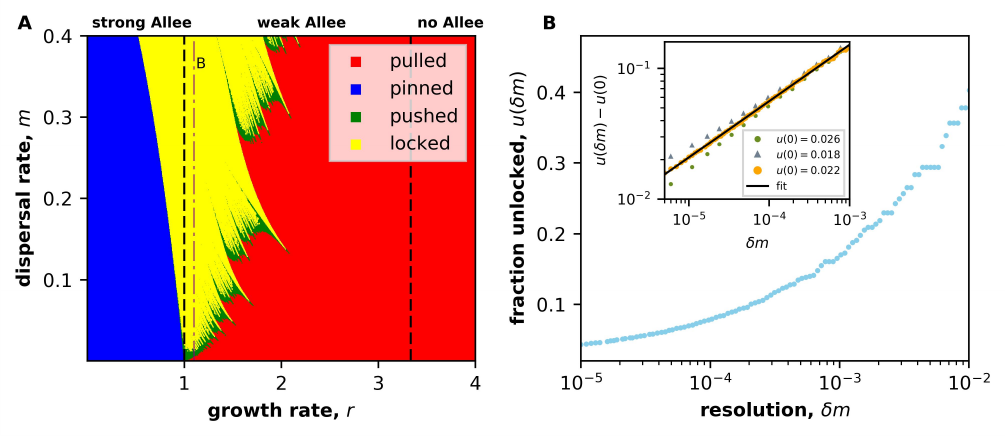}} 
\caption{\textbf{Pinned, locked, pushed, and pulled invasions in DSDT models.} \textbf{(A)} shows the location of all four expansion classes in the parameter space. This phase diagram was obtained for the piecewise-linear model with~$c^*= 0.3 $ and~$K= 1.0$, and we varied~$r$ and~$m$ in steps of~$\delta r = 0.004$ and~$\delta m = 0.0004$. Waves were labeled as pulled if their velocity was less or equal to the prediction of Eq.~(\ref{eq:v_min}) with a tolerance of~$10^{-5}$ to account for the finite precision of our simulations. Waves were classified as locked if the measured velocity differed by less than~$10^{-5}$ from one of the neighboring simulations. \textbf{(B)} argues that the results in A are not an artifact of finite resolution in the parameter space. A specific value of~$r=1.1$ (dash-dotted line in A) was chosen to examine how the fraction of unlocked waves,~$u$, depends on the resolution,~$\delta m$, at which~$v(m)$ were obtained. The bending of the~$u(\delta m)$ curve suggests that~$u(0)>0$, i.e. unlocked waves occupy a region with nonzero measure in the parameter space. To better estimate~$u(0)$, we performed the following fit:~$u(\delta m)\approx u(0) + A (\delta m)^\beta$; see~\citep{arnold:scaling, fat_fractal:prl, eykholt:scaling}. The inset illustrates the quality of this fit for~$u(0)= 0.0221 \pm 0.0001$ and~$\beta=0.430 \pm .001$~(yellow dots), which were obtained by maximizing $R^2$. Green and grey symbols demonstrate that clear deviations from the power-law dependence are observed for~$u(0)$ slightly different from the best estimate. Importantly,~$u(0)$ is significantly greater than the fraction of pulled waves, which is~$0.015$ for~$r=1.1$ and~$m\in(0,0.4)$. Thus, some waves must be pushed because they are neither locked nor pulled. In panel B simulations were run at $m$ in steps of~$\delta m = 10^{-6}$, hence locked waves were classified with a velocity tolerance~$5 \times 10^{-7}$. The fit was performed in the interval~$(5 \times 10^{-6},10^{-3} )$ and error bars were obtained from bootstrapping. } 
\label{fig:phase_diagram}
%\end{center}  
\end{figure}

We then asked whether locked and pulled waves are the only two classes of expansions in DSDT models or whether DSDT models also support unlocked waves that are pushed rather than pulled. To answer this question, we obtained the phase diagram of invasion velocities in the piecewise-linear growth model. Specifically, we computed~$v$ from simulations with~$K= 1.0$,~$c^* =0.3$, and~$r$ and~$m$ varied by~$\delta r = 0.004$ and~$\delta m = 0.0004$ to completely tile the region of~$(r,m)\in(0, 4)\times(0,0.4)$. The results are shown in Fig.~\ref{fig:phase_diagram}A. Waves were labeled as pulled if their velocity differed from Eq.~(\ref{eq:v_min}) by less than~$10^{-5}$. Locked waves were defined as waves whose velocity differed by less than~$10^{-5}$ from the velocities in at least one of the neighboring simulations, i.e. simulations with parameters different by~$\delta r$ or~$\delta m$. Locked invasions with~$v< 10^{-4} $ were labeled as pinned. Our main result was that some parameter combinations for which the invasions could not be classified as either pulled or locked. That is their velocities appeared to change continuously with model parameters, yet the velocities were greater than the expectation for pulled waves. In analogy with CSCT models, we labeled such invasions as pushed.

One might wonder whether the existence of pushed waves in Fig.~\ref{fig:phase_diagram}A could be attributed to finite~$\delta r$ and~$\delta m$ in our simulations. Indeed, velocities plateaus smaller than~$\delta r$ and~$\delta m$ are missed by the procedure described above. To exclude the artifacts due to finite resolution, we considered a slice of the phase diagram at~$r= 1.10$ and examined how the fraction of unlocked waves~$u$ depends on~$\delta m$. We found that~$u(\delta m)$ approaches as finite value~$u(0)$ as~$\delta m \to 0$; see Fig.~\ref{fig:phase_diagram}B. Thus, not all expansions are locked at~$r=1.10$. Following a similar analyses of mode locking in nonlinear oscillators~\citep{arnold:scaling, fat_fractal:prl, eykholt:scaling}, we improved the estimate of~$u(0)$ by fitting~$u(\delta m)$ to a power-law functional form that is expected on the theoretical grounds:~$u(\delta m)\approx u(0) + A \delta m^\beta$. The fit with the highest~$R^2$ was obtained for~$\beta = 0.430 \pm .001$ and~$u(0)= 0.0221 \pm 0.0001$. Importantly, the inferred fraction of unlocked waves is significantly greater than the fraction of pulled waves. The latter equals~$0.015$ for the parameters studied, $r=1.1$ and~$m\in(0,0.4)$. Thus, some waves must be pushed because they are neither locked nor pulled. Further evidence for the existence of pushed waves is provided by Fig.~\ref{fig:dispersal}B, where the fraction of pushed expansions is much larger.

The transitions between different modes of propagation in DSDT models is more complex than in other models. For CSCT model, the transition between pulled and pushed waves occurs at a specific strength of the Allee effect and is independent of~$m$. In contrast, the phase boundary depends on both~$r$ and~$m$ in DSDT models. Moreover, the boundaries between locked, pushed, and pulled waves have fractal geometry due to an infinite number of velocity plateaus, each occupying a distinct region in the parameter space. The regions themselves have a smooth boundary and are known as Arnold tongues~\citep{arnold:1983_translated, arnold:scaling}. Pinned waves correspond to a single tongue with~$v=0$ and, therefore, are separated by a smooth boundary from other types of expansion dynamics.

\subsection{Response to perturbations}
To better understand the differences between pulled, pushed, and locked invasions, we examined how they respond to perturbations. In particular, we were interested in how the effect of a small perturbation decays in time and whether this effect vanishes at long times or not. To this end, we simulated invasion profiles until they reached a steady state and then perturbed the population density by replacing~$c$ with~$c + \epsilon c(1-c)$. We then examined the difference in~$c(t,x)$ for perturbed and unperturbed copies of the system.

\begin{figure}[!h]
%\begin{center}
\centerline{\includegraphics[width=0.5\linewidth]{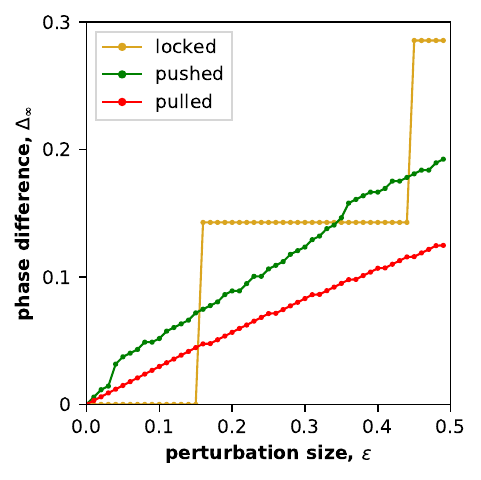}}
\caption{\textbf{Response to perturbation is different for pulled, pushed and locked waves.} The phase difference, $\Delta_{\infty}$, between the unperturbed and perturbed profiles after $10^{5}$ time steps as a function of the perturbation strength $\epsilon$ is shown. We define~$\Delta_{\infty}$ as the number by which the profile of the perturbed wave needs to be shifted back to match the profile of the unperturbed wave. For locked expansions,~$\Delta_{\infty}$ takes discrete values including~$\Delta_{\infty}=0$ for weak perturbation. For pushed and pulled waves,~$\Delta_{\infty}>0$ provided~$\epsilon>0$ and the dependence is continuous. Note that we observe slight undulations in~$\Delta_{\infty}(\epsilon)$ for pushed, but not pulled waves. Parameters for the plots are  $c^{*} = 0.3$, $K=1.0$ for the piecewise-linear growth function. For locked, pushed and pulled the growth rates are, $r=1.096, 1.52, 3.33$ and the migration rates are $m=0.068, 0.1432, 0.20$ respectively.} 
\label{fig:perturbations}
%\end{center}  
\end{figure}

We expected that a perturbation of pulled and pushed fronts should lead to a nonzero shift of the front profile. This expectation follows from the following argument. The movement of unlocked fronts is described by a continuous function~$c(x-vt)$, which is a solution of Eq.~(\ref{eq:dsdt}). Because~$c(x-vt)$ is continuous and~Eq.~(\ref{eq:dsdt}) is translationally invariant,~$c(x-vt+\mathrm{const})$ is also a solution that must describe the front propagation from different initial conditions. Given that perturbed and unperturbed systems have slightly different initial conditions their long-time behavior should in general be different by a shift along the~$x$-axis. To determine the shift, first we fit the perturbed and unperturbed profiles using cubic splines, and then found the shift along the x-axis which minimized the differences between the profiles. We called this quantity the phase difference,~$\Delta_{\infty}$, between the asymptotic profiles. In simulations, we waited for $10^{5}$ generations after the perturbation before fitting the splines by using profiles from $10^{3}$ generations. Our simulations, shown in Fig.~\ref{fig:perturbations}, confirmed a non-zero phase difference for pushed and pulled waves for small perturbations, but showed a different behavior for locked invasions. Locked fronts returned to their unperturbed positions provided the perturbation did not exceed a certain threshold; otherwise, the position of the front was shifted by an integer number of patches. This difference between locked and unlocked waves is not surprising. The former has a discrete number of distinct steady-state profiles in which the front can settle. Each of these steady-state fronts has a nonzero basin of attraction, so a perturbation must exceed a certain threshold to force the invasion from one steady state into another. 

\begin{figure}[!h]
%\begin{center}
\centerline{\includegraphics[width=\linewidth]{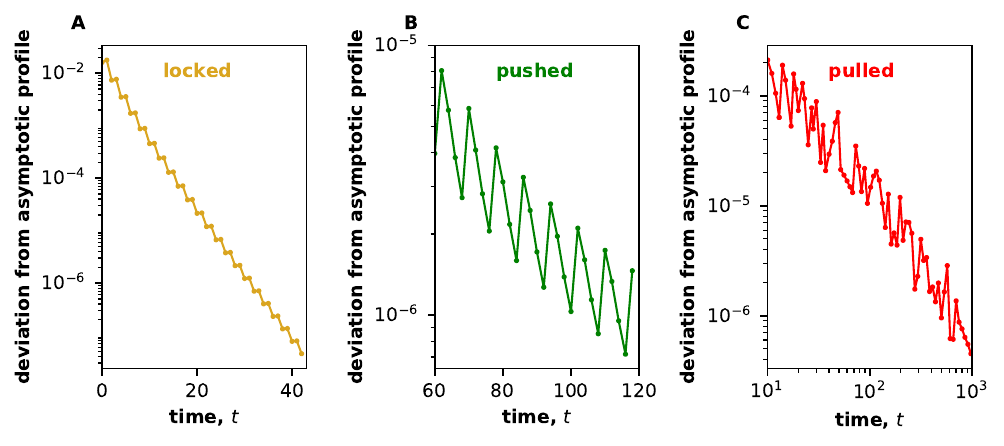}}
\caption{\textbf{Locked, pushed and pulled invasions have different relaxations after perturbation.} The deviation between the density profile of the perturbed wave and asymptotic profile in time for perturbation strength $\epsilon=0.75$. \textbf{(A,B)} Locked and pushed waves showed an exponential decay, which is a straight line on the log-linear plot. \textbf{(C)} Deviation for pulled wave decayed as~$1/t$ which is a straight line on a log-log plot. All panels used the piecewise-linear growth map with $c^{*}=0.3$, and $K=1.0$. The growth and dispersal rates were $r=1.1, m=0.38$ for the locked wave, $r=1.52, m=0.1432$ for pushed waves and $r=3.33, m=0.20$ for the pulled wave.  }
\label{fig:relaxation}
%\end{center}  
\end{figure}

There are also differences in how fronts approach the long-time steady state following a perturbation. Exponential relaxation is expected for locked waves, since each of the discrete profiles is a stable fixed point of the dynamics. In contrast, pulled waves are expected to show a much slower relaxation with perturbations decaying as~$1/t$. The reasons for this non-exponential relaxation is that the linearized Eq.~(\ref{eq:dsdt}) admits solutions with a continuum of possible velocities and, therefore, its spectrum does not have a gap~\citep{saarloos:review, birzu:semipushed}. Both of these predictions are confirmed by our simulations shown in Fig.~\ref{fig:relaxation}. We also found that perturbations decay exponentially in time for pushed waves. This observation is consistent with the results for CSCT models and is related to the existence of a unique velocity, and, therefore, a spectral gap for pushed waves~\citep{saarloos:review, kessler:velocity_cutoff}.  

To determine the relaxation to perturbations, we compared the profile of the wave just after perturbation to the asymptotic profile. The asymptotic profile was determined by waiting $10^{5}$ time steps after perturbation and fitting the following $10^{3}$ profiles using a cubic spline. The profiles at each time point after the perturbation were shifted along the~$x$-axis to find the minimum deviation from the asymptotic profile. This minimum deviation shows characteristic decay which depends on wave being pulled, pushed or locked in fig.~\ref{fig:relaxation}.

\subsection{Locked invasions are robust to fluctuations}

Since locked waves return to exactly the same steady state following perturbations, they should be robust to demographic and environmental fluctuations. To establish the limits of this robustness, we modified the piecewise-linear growth model and Eq.~(\ref{eq:dsdt}) to include demographic noise, temporal fluctuations of the environment, or spatial heterogeneity of the habitat.  

Demographic fluctuations were described by
\begin{equation}
c_{t+1,x} =  \textsf{Bi} \left[f\left( \frac{m}{2}c_{t,x-1} + (1-m)c_{t,x} + \frac{m}{2}c_{t,x+1} \right) ,N \right]
\label{eq:dsdt_demographic}
\end{equation}

\noindent where $\textsf{Bi} [p,N] $ refers to binomial sampling from $N$ with probability $p$ . Temporal fluctuations were modeled by drawing~$m$ from a uniform distribution from~$m-\Delta m/2$ to~$m+\Delta m/2$ at each time step in Eq.~(\ref{eq:dsdt}). Spatial heterogeneity was included by using a different~$r$ in each patch, which was drawn from a uniform distribution between~$r-\Delta r/2$ and~$r+\Delta r/2$. We measured the velocity in each scenario by a linear fit to the sum of the concentration normalized by the carrying capacity for a single run.

\begin{figure}[!htb]
\includegraphics[width=\linewidth]{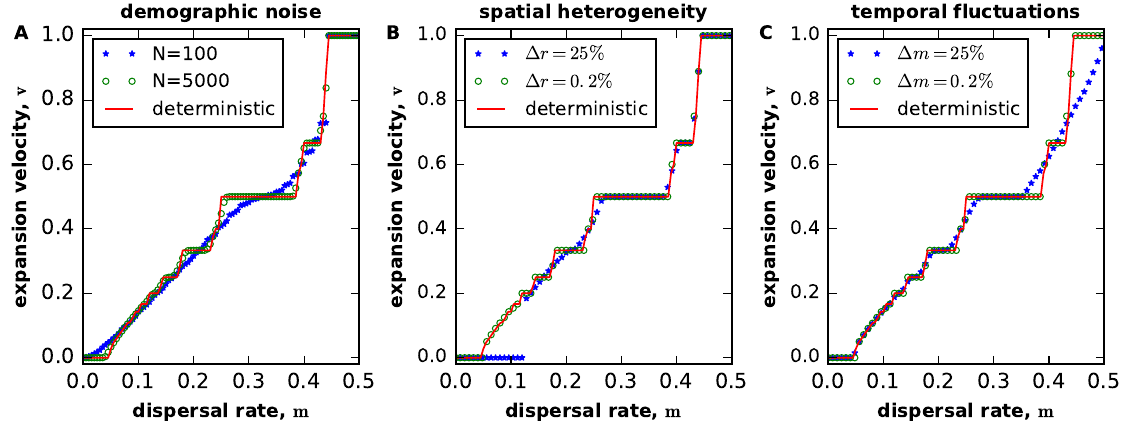}
\caption{\textbf{Velocity locking is robust to demographic, temporal and spatial fluctuations.} \textbf{(A)} Demographic fluctuations in a population of size~$N$ is modeled via binomial sampling from~$N$. Velocity locking is unaffected for sufficiently large population sizes. \textbf{(B}) Locking is robust to spatial variation of growth rates. The growth rates at each point in space were drawn from a uniform distribution between~$r-\Delta r/2$. and~$r+\Delta r/2$. Large fluctuations can lead to extended regions where the growth rate is small, increasing the width of the pinned region. \textbf{(C)} Locking is robust to temporal variation of migration. Dispersal rates at each time step were sampled from a uniform distribution between~$m-\Delta m/2$ and~$m+\Delta m/2$ . Parameters used in the figure are ~$r=0.93$,~$c^*=0.22$. } 
\label{fig:fluctuations}
\end{figure}

For all three scenarios, we observed that velocity locking is extremely robust to fluctuations~(Fig.~\ref{fig:fluctuations}). Although small plateaus are progressively washed out by stronger fluctuations, large plateaus remain clearly visible even when fluctuations are about 25\% of the mean. Thus, velocity locking should be readily observable in laboratory and potentially natural populations.

\subsection{Velocity locking suppresses front diffusion}
Given that velocity plateaus remain even in the presence of strong fluctuations, we decided to examine how velocity locking affects stochastic properties of invasion fronts. The consequences of demographic and environmental noise are understood relatively well in CSCT models~\citep{birzu:semipushed, mikhailov:diffusion, rocco:diffusion, meerson:velocity_fluctuations, panja:review}. The main effects are fluctuations of the front shape and diffusive wandering of the front position, which could be defined in a number of ways, for example, as the integral of population density normalized by the carrying capacity. For an ensemble of independent simulations, the mean position of the front grows linearly in time as in deterministic simulations. The variance of the front position also grows linearly in time as if the front is performing an unbiased random walk relative to its expected position. The effective diffusion constant of the front,~$D_f$, can then be obtained fitting the linear increase of the variance to~$2D_{f}t$ simulations;~$D_f$ can also be computed analytically~\citep{birzu:semipushed, mikhailov:diffusion, rocco:diffusion, meerson:velocity_fluctuations, panja:review}.  

\begin{figure}[!htb]
\includegraphics[width=0.5\linewidth]{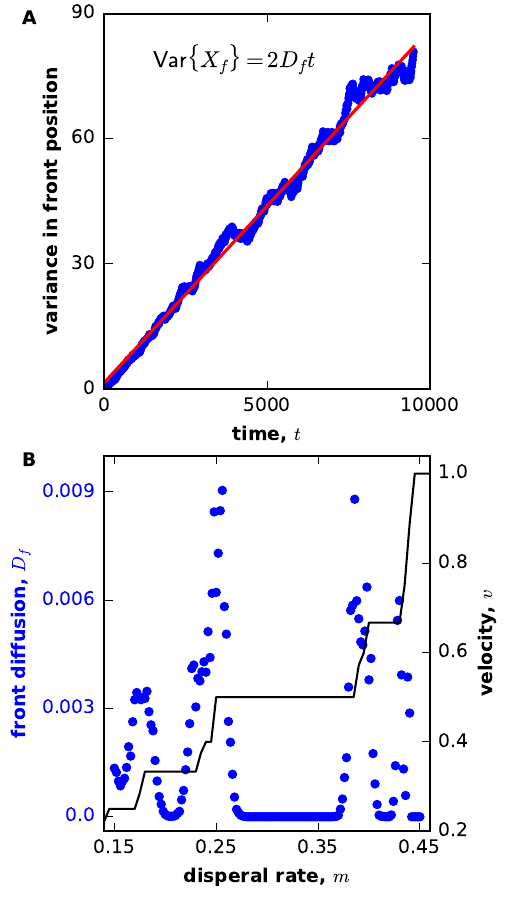}
\caption{\textbf{Dramatic changes in front diffusion between locked and unlocked regimes.} \textbf{(A)}~shows that the variance of the front position increases linearly in time and can therefore be used to define effective diffusion constant of the front,~$D_f$. \textbf{(B)}~The dependence of~$D_f$ indicates major differences between locked and unlocked regimes. Inside velocity plateaus~$D_f=0$, but it is nonzero in unlocked regions. The diffusion constant is especially large near plateau boundaries, where~$v(m)$ changes rapidly. For convenience,~$v(m)$ is plotted in thin black line. Here, we simulated the piecewise-linear model with demographic noise:~$r=0.93$,~$c^*=0.22, N=2000$. To determine each value of~$D_f$, we ran simulations for~$ 9500 $ generations and averaged over~$100$ independent runs.} 
\label{fig:front_diffusion}
\end{figure}

We applied this procedure of estimating~$D_f$ to DSDT models where we defined the position of the front as the sum of the population density normalized by the carrying capacity. Figure~\ref{fig:front_diffusion}A shows that noisy invasions in DSDT models also exhibit diffusive front wandering and, therefore, can be assigned an effective diffusion constant.  We examined how this diffusion constant depends on velocity locking by measuring~$D_f$ as a function of~$m$ for parameter values that result in both locked and unlocked fronts~(Fig.~\ref{fig:front_diffusion}B). We found that~$D_f$ vanishes on velocities plateaus, but shows very large peaks immediately outside the plateaus. These findings parallel the behavior found in mode locked systems~\citep{rubi:locking_noise}, and can be explained by the dynamics without noise. We showed above, see Fig.~\ref{fig:perturbations}, that, for locked invasions, small perturbations are quickly forgotten and, ultimately, have no effect on the front position. Therefore,~$D_f$ must be zero in this case. For pulled and pushed invasions, however, any perturbations leads to a small shift in the asymptotic position of the front, so repeated perturbations due to demographic or environmental fluctuations result in a random walk with~$D_f>0$. The effective diffusion constant is particularly large near the boundaries of the plateaus because~$v(m)$ changes rapidly in these regions and, therefore, perturbations have a much larger effect on front motion. 

The vanishing of~$D_f$ for locked invasions provides a convenient way to detect velocity locking in situations where one cannot modify the model parameters and confirm the existence of a velocity plateau. Zero diffusion could also be beneficial in technological applications that require coherence or reproducibility. In such case, one might prefer to operate the system at one of the velocity plateaus. On the other hand, the anomalously large~$D_f$ near plateau boundaries could explain extreme variability of some invasions or could be used to amplify variability in situations where it is beneficial.       

\subsection{Locked waves due to positive density-dependent dispersal} 
Mode locking requires nonlinear dynamics. So far, we focused on the Allee effect as the source of this nonlinearity. In the context of range expansions, density-dependent dispersal could provide an alternative mechanism. To test this hypothesis, we modified Eq.~(\ref{eq:dsdt}) as follows

\begin{equation}
    c_{t+1,x} = f\left[ \frac{m(c)}{2}c_{t,x-1} + \left(1-m(c)\right)c_{t,x} + \frac{m(c)}{2}c_{t,x+1}   \right],
\label{eq:dispersal}
\end{equation}
\noindent and used a simple linear dependence of~$m$ on~$c$:

\begin{equation}
m(c)=m_0 + m_1 c
\label{eq:ddd}
\end{equation}

\noindent For~$f(c)$, we used the piecewise-linear model without an Allee effect~($rc^*=K$).

The linear dependence of~$m$ on~$c$~(Eq.~\ref{eq:ddd}) was observed in some species~\citep{morisita:linear_diffusion}  and has been frequently used in mathematical modeling~\citep{murray:mathematical_biology, kawasaki:density_dependent_diffusion}. For~$m_1<0$, the dispersal is highest at the edge of the front, and the invasions are pulled. For~$m_1>0$, the density-dependence is positive, and the linear theory of pulled waves may not apply. 

Indeed, the transition from pulled to pushed expansions has been recently demonstrated in a CSCT model with density-dependent dispersal~\citep{kawasaki:density_dependent_diffusion}, which can be specified by the following partial differential equation

\begin{equation}
\frac{\partial c}{\partial t}=\frac{1}{2}\frac{\partial^2}{\partial x^2}[m(c)c]+g(c).
\end{equation}

\noindent For logistic growth and linear~$m(c)$, \citep{kawasaki:density_dependent_diffusion} obtained exact expressions for the velocity and front shape of pushed waves:

\begin{equation}
	v = \sqrt{\frac{m_1 r K}{2}} + m_0 \sqrt{\frac{r}{2m_1 K}}.
	\label{eq:v_density_dependent_diffusion}
\end{equation}

\noindent This formula is applicable for~$ m_1K>m_0$. For lower values of~$m_1$, the expansion is pulled, and its velocity is given by~$v=\sqrt{2m_0 r}$. We note that the results in \citep{kawasaki:density_dependent_diffusion} are based on a general mapping between equations that determine expansion velocity in models with density-dependent growth and density-dependent dispersal. This mapping was first reported in \citep{hadeler:mapping}.

\begin{figure}[!h]
%\begin{center}
\centerline{\includegraphics[width=\linewidth]{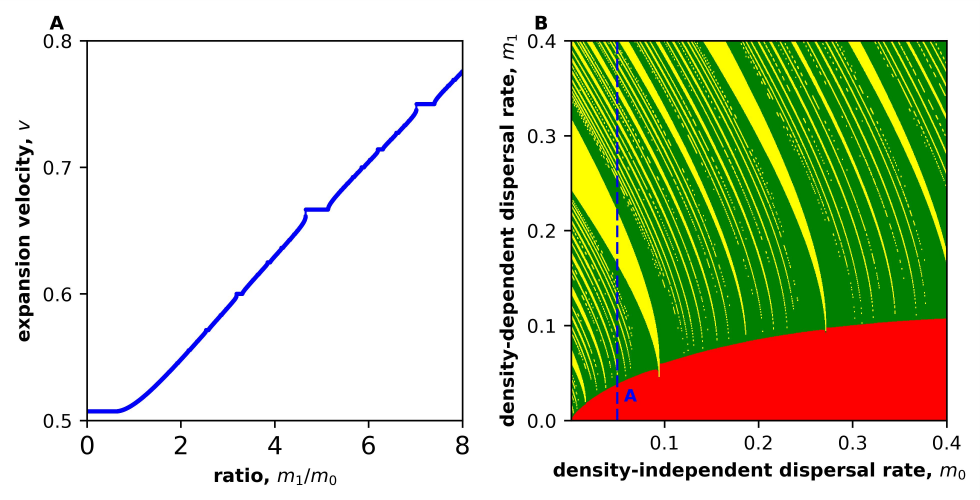}}
\caption{\textbf{Velocity locking can result from positive density-dependent dispersal.} \textbf{(A)}~shows how expansion velocity depends on~$m_1/m_0$, the ratio of density-dependent to density-independent components of dispersal in the model defined after Eq.~(\ref{eq:dispersal}). For small rations~$v(m)$ looks smooth indicating no velocity locking, but large~$m_1/m_0$ show clear velocity plateaus. In these simulations, $m_{0}= 0.05 $ was kept constant as $m_{1}$ is varied; the growth parameters were chosen not to have an Allee effect:~$r=3.33$,~$c^*= 0.3 $,~and~$K=1.0$.  \textbf{(B)}~shows the regions of pulled, pushed, and locked waves in the~$m_0-m_1$ plane. Note that there are no pinned waves because they require a strong Allee effect. The color scheme and numerical procedures are the same as in Fig.~\ref{fig:phase_diagram}A; the growth parameters are the same as in panel A.} 
\label{fig:dispersal}
%\end{center}  
\end{figure}

In DSDT model, increasing the ratio of~$m_1/m_0$ also shows a clear transition form pulled to pushed and locked expansions. For pulled waves,~$v(m_1/m_0)$ is smooth while expansions at higher~$m_1/m_0$ exhibit velocity plateaus~(Fig.~\ref{fig:dispersal}A). The phase diagram in the~$m_0-m_1$ space is qualitative similar to that in~$r-m$ space, but pushed waves are more prevalent in the~$m_0-m_1$ space presumably because both~$f(c)$ and~$m(c)$ are continuous in Fig.~\ref{fig:dispersal}A, but~$f(c)$ is discontinuous in Fig.~\ref{fig:phase_diagram}.

Density-dependent dispersal has been described in many species, and it can also be easily engineered in microbes~\citep{dispersal:review,hwa:engineered_dispersal,dispersal:sheep}. Therefore, nonlinearities in dispersal could provide an alternative route to velocity locking in populations of living organisms. 

\subsection{Evolution in locked invasions} 
Given that small changes in model parameters do not affect the velocities of locked fronts, we were interested to determine whether velocity locking affects species evolution. Specifically, we asked whether velocity locking eliminates the selection for higher dispersal or growth, which has been repeatedly observed in many species~\citep{ditmarsch:swarming, phillips:toad_acceleration, phillips:toad-theory, korolev:sectors}. To this purpose, we considered the competition between two genotypes with different dispersal rates~(the results for different growth rates were similar). The density of the genotypes are labeled~$c^{a}$, where~$a=1$ for the resident and~$a=2$ for the mutant. The update is divided into an intermediate migration step followed by a growth step, given by:
\begin{equation*}
\begin{aligned}
    c^{a}_{t+\tau, x} & = \frac{m^{a}}{2}c^{a}_{t,x+1} +  (1-m^{a})c^{a}_{t,x} + \frac{m^{a}}{2}c^{a}_{t,x-1}
    \\
    c^{a}_{t+1,x} & = \frac{c^{a}_{t+\tau,x}} {\sum_{a} c^{a}_{t+\tau,x} } g\left( \sum_{a} c^{a}_{t+\tau,x} \right)
\end{aligned}
\label{eq:evolution}
\end{equation*}
\noindent where~$t+\tau$ denotes the intermediate step between $t$ and $t+1$, and $f(c)$ is the piecewise-linear growth map. The second equation implements growth in density while  preserving the the ratio of resident and mutant from the migration step. 

\begin{figure}[!h]
%\begin{center}
\centerline{\includegraphics[width=0.5\linewidth]{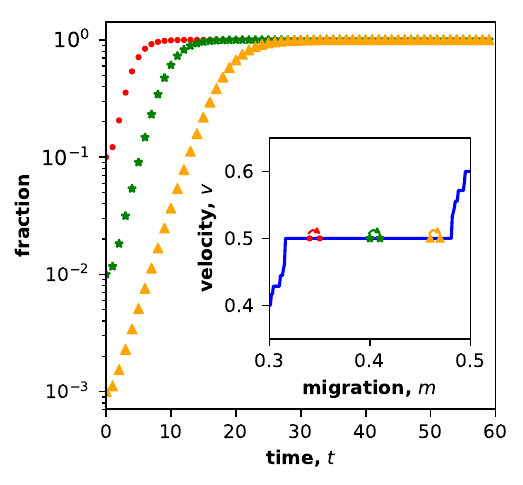}}
\caption{\textbf{Invasion of mutant with higher migration rate but same propagating velocity.} Three sets of resident and invaders are chosen such that they have the same velocity as they are part of the same Arnold tongue. The migration rate of the mutant is $0.01$ greater than the resident. We used different initial fractions of the invaders, namely $(0.1, 0.01, 0.001)$ to improve readability of the figures. Parameters of the simulation are $r=1.1$, $c^{*}=0.3$ and $K=1.0$ } 
\label{fig:evolution}
%\end{center}  
\end{figure}

We found that mutants with different dispersal rate can take over the population even though the mutant and the resident have identical invasion velocities~(Fig.~\ref{fig:evolution}). Similar to previous findings for CSCT models~\citep{korolev:arrest}, selection for both faster and slower dispersal was possible depending on the strength of the Allee effect. We therefore conclude that velocity locking does not arrest evolutionary dynamics in the expanding population.

\subsection{Velocity locking in two spatial dimensions} 
Since many range expansions occur in two rather than in one spatial dimensions, we sought to determine how spatial dimensionality affects velocity locking. To answer this question, we generalized Eq.~(\ref{eq:dsdt}) to two spatial dimensions:

\begin{equation}
c_{t+1,(x,y)} = f\left((1-m)c_{t,(x,y)} + \frac{m}{4}(c_{t,(x+1,y)} + c_{t,(x-1,y)} + c_{t,(x,y+1)} +c_{t,(x,y-1)} )\right).
\end{equation}

\begin{figure}[!h]
%\begin{center}
\centerline{\includegraphics[width=\linewidth]{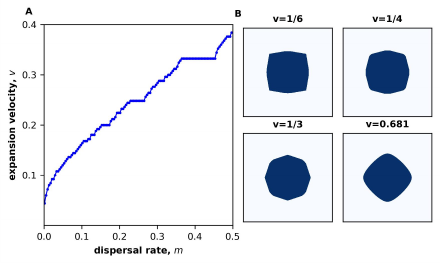}}
\caption{\textbf{Velocity locking occurs in two dimensions.} \textbf{(A)} The velocity as a function of $m_{0}$ shows distinct plateaus for $r=1.2$. Velocity was measured by fitting the position of front. \textbf{(B)} In two dimensions, the colony takes different shapes depending on the velocity of the locked wave. Pulled waves form a diamond shape. The shapes of the colonies at the end is independent of the initial shape. In Panel A, the piecewise-linear growth function with $r=1.2$, $c^{*}=0.3$, $K=1.0$ is used. For all four runs, the common parameters are $m=0.2$, $K=1.0$, $c^{*}=0.3$. In Panel B, the growth rates, $r=0.95, 1.15, 1.35$ and $3.33$ corresponds to the velocities, $v=1/6, 1/4, 1/3$ and $0.681$. A static box of 1000x1000 demes was used for all runs in two dimensions. }   
\label{fig:2d}
%\end{center}  
\end{figure}

The results of two-dimensional simulations are shown in Fig.~\ref{fig:2d}. We found that velocity plateaus are still present in two spatial dimensions; in addition, the fronts of the expansions assume characteristic geometric shapes that depend on the magnitude of the expansion velocity.

\subsection{Velocity locking in CSCT models with spatial and temporal periodicity} 
Although velocity locking arises most naturally in DSDT models, it is the spatio-temporal periodicity rather than discreteness that is required for velocity locking. To demonstrate this explicitly, we simulated the following CSCT model with the growth~$g(t,x)$, dispersal~$D(t,x)$, and death~$j(t,x)$ rates varying periodically in time and space:

\begin{figure}[!h]
\includegraphics[width=0.5\linewidth]{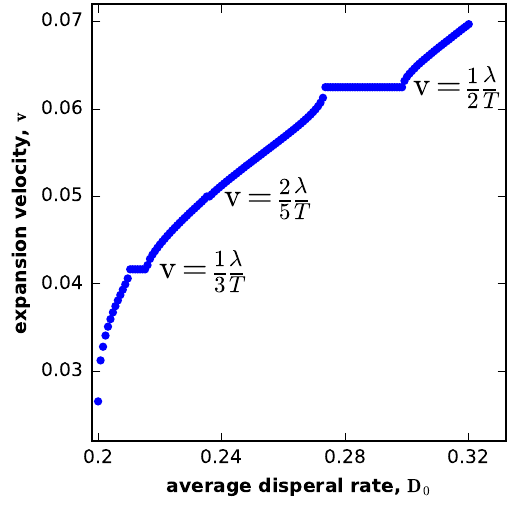}
\caption{\textbf{Velocity locking in CSCT models with spatial and temporal periodicity.} The figure shows three distinct velocity plateau in a fully continuous reaction-diffusion model. The dispersal, growth, and death rates varied in time and space according to Eqs.~(\ref{eq:continuous_dgj}) with temporal period $T=160$ and spatial period~$\lambda=20$. The fundamental velocity imposed by this spatio-temporal structure is then~$\frac{\lambda}{T}=0.125$. For all locked invasions, the expansion velocities equaled a rational number times the fundamental velocity. Here, $D_1=0.2,~g_0=1.0,~j_1=0.32,~c_{a}=-0.2$, and we varied $D_0$. Simulations were run for~$\Delta t = 0.0625$, and~$\Delta x$ was chosen to ensure a sufficiently fine discretization.} 
\label{fig:continuous}
\end{figure}

\begin{equation}
\frac{\partial c}{\partial t}= \frac{\partial }{\partial x} \left( D(t,x) \frac{\partial c}{\partial x} \right) + [g(t,x)-j(t,x)]c.
\label{eq:continuous}
\end{equation}

\noindent To specify the periodic variation, we used a simple cosine with an offset:

\begin{equation}
\begin{aligned}
D(t,x) =&  D_0  \left[1 + \cos \left(2 \pi \frac{t}{T}  \right)  \right] +  D_1  \left[1 - \cos \left(2 \pi \frac{t}{T}  \right)  \right]  \left[1 + \cos \left(2 \pi \frac{x}{\lambda} \right)\right], \\
g(t,x) =&  \left[1 + \cos \left(2 \pi \frac{t}{T}  \right)  \right] g_0  c(1-c)\left(c-c_{a}\right), \\
j(t,x) =&   j_1  \left[1 - \cos \left(2 \pi \frac{t}{T}  \right)  \right] \left[1 + \cos \left(2 \pi \frac{x}{\lambda} \right)\right],\\
\end{aligned}
\label{eq:continuous_dgj}
\end{equation}

\noindent where~$T$ and~$\lambda$ are temporal and spatial periods, and ~$D_1$ and~$j_1$ set the magnitude of the spatial variation. Note that the growth rate is identical to that in Eq.~(\ref{eq:cubic}) with~$K=1$ except for the temporal variation. The intuition behind Eqs.~(\ref{eq:continuous_dgj}) is as follows. For the first half of the season, the growth rate is above average, and, for the second half, the death rate is above average. The death and dispersal rates also vary spatially, with higher dispersal rates in the regions with higher death rates. 

Equation~(\ref{eq:continuous}) was solved numerically as described in Methods. Velocity was computed by fitting the front position to a linear function of time. The front position was defined as the furthest site with population density greater than half the average bulk density at the time. The velocity of invasion is shown in Fig.~\ref{fig:continuous} as a function of the average dispersal rate~$D_0$. There are clear velocity plateaus at rational multiples of the fundamental velocity,~$\frac{\lambda}{T}$. Thus, velocity locking can occur in CSCT models provided there are periodic modulations of the population dynamics in both time and space.

%**************************************************************************
% Discussion
%**************************************************************************
\section{Discussion}

In this paper, we examined the properties of range expansions in fragmented habitats with seasonal growth. Both ecological conditions are quite common in natural and laboratory populations~\citep{keitt:pinning, zhang:death_galaxy, fisman:seasonality, fahrig:fragmentation, wilson:coexistence, kefi:desert, abraham:patchy_plankton, nelson:flow_prl_2012, datta:wave_splitting, gandhi:pulled_pushed, dai:nature, zhang:death_galaxy, active_matter:complex_environment, jorn:vortex_lattice, tsimring:calcium_pinning, paoletti:epl_front_locking}; therefore, understanding their effect could have important applications in biotechnology and ecosystem management. This is especially true for systems with liquid-handling robots or microfluidic devices, regular population arrangements found in agricultural settings, and for species shifting towards more fragmented and seasonal habitats in response to raising temperatures. Invasions under such conditions are better described by models with discrete space and time rather than partial differential equations~\citep{mistro:complexity_allee, lewis:chamomile}. Discrete models also arise when continuous models are solved on a computer. Although continuous behavior is typically assured in the limit of infinitely fine discretization,\footnote{The transition from DSDT to CSCT models and the associated smoothing of velocity plateaus were studied in Ref.~\citep{carretero:locking_pre}} large discretization steps are often chosen in practice because one is willing to sacrifice accuracy for computational efficiency. The key question is then whether the difference between continuous and discrete models is only quantitative or whether there are qualitative differences between the alternative approaches to model population dynamics.
 
When both space and time are discrete, range expansions can fundamentally differ from the predictions of continuous models. In particular, invasions can proceed in a step-like or pulsed fashion with population densities at the front assuming a discrete set of values that repeat periodically. Such invasions have been observed in the field although they have been attributed to other factors~\citep{johnson:allee_gypsy,sullivan:pulsed_invasions}. The velocities of pulsed expansions are locked and completely insensitive to moderate variations in the rates of dispersal and growth. However, when parameter change exceeds a certain threshold, expansion velocity and the periodicity of front pulsations change discontinuously. These dynamics produce a characteristic pattern of plateaus on the plot of a response variable~(velocity) vs. a control variable~(migration) known as the Devil's staircase in physics~\citep{bak:devils_staircase}; see Fig.~\ref{fig:locking_common}.

Why does velocity locking occur only in DSDT models and not for pulled waves? The answer lies in the translational symmetry of the traveling wave solutions. Consider an expansion that has reached a steady state and expands at constant velocity with a time-invariant density profile in the reference frame co-moving with the expansion. When space is continuous, a shift of the density profile by any distance will result in a density profile that is also a traveling wave solution of the underlying mathematical model. In consequence, there is no restoring force that would oppose a spatial translation of the invasion front, and any perturbation would result in a nonzero shift in the front position. Repeated perturbations would then modify the invasion rate, and, more generally, we expect that changes in the model parameters would result in nonzero changes in the expansion velocity. Thus, models with continuous space cannot exhibit velocity locking.

Invasions in DSCT models also possess translational symmetry because two consecutive density profiles separated by any time are both solutions of the underlying mathematical model. Since time is continuous, there is again a continuous family of solutions, which are also related by spatial translations since density profiles at different times are simply shifted relative to each other. The only exception are pinned invasions because a temporal translation does not result in a distinct density profile when the velocity is zero. As a result, DSCT models exhibit velocity pinning, but not locking for exactly the same reasons as the continuous space models. 

For pulled waves, velocity locking does not occur because of another continuous symmetry unrelated to spatial and temporal continuity. The velocity of pulled invasions is determined by the dynamics at the expansion edge, where the population density is small and all nonlinear terms can be neglected. Since linear equations are invariant under multiplication by any positive number, there is a continuous family of solutions, which can be interpreted as temporal or spatial translations. Using the same arguments as for CSCT models, we then conclude that pulled expansions should never experience velocity locking.  

This symmetry based analysis predicts that models that do not exhibit velocity locking should have continuous profiles in the comoving reference frame. This prediction agrees with the known results and our simulations. When space is continuous, the density profile is clearly a continuous function, which translates in space as the expansion proceeds~\citep{murray:mathematical_biology, saarloos:review, kot:csdt, wang:csdt}. Although the density profile at any given time assumes only a discrete set of values in DSCT models, the density profile is nevertheless continuous in the comoving reference frame when density profiles at different times are shifted relative to each other by the distance that they have traveled~\citep{kot:csdt, wang:csdt}. Similarly, pulled and pushed waves in DSDT models have a continuous density profile in the comoving reference time despite the fact that both space and time are discrete~(Fig.~\ref{fig:understanding_locking}A and Appendix). In sharp contrast, the density profile during velocity locking is clearly discrete even in the comoving reference frame~(Fig.~\ref{fig:pulled_works}).

The remaining question is why velocity locking occurs in DSDT models with significant nonlinearities in dispersal or growth. An intuitive argument was proposed in Ref.~\citep{carretero:locking_pre} that noticed that the infinite dimensional system of coupled difference equations in DSDT models can be reduced to a single difference equation. This dimensional reduction is based on time scale separation between the dynamics of front shape and position, which occurs because perturbations decay exponentially fast in invasions that are not pulled~(Fig.~\ref{fig:relaxation}). Thus, the position of the front~$X_f$ can be described by the following map

\begin{equation}
\label{eq:circle_map}
X_f(t+1)=F(X_f(t)).
\end{equation}

\noindent The function~$F$ is necessarily periodic, so the dynamics can be restricted to~$X_f\in(0,1)$ for DSDT models or to~$X_f\in(0,\lambda)$ for periodic CSCT models~(in this case one considers~$X_f$ at times~$t$ and~$t+T$). Because of this periodicity~$F$ is often referred to as a circle map. 

Mode locking in circle maps is well understood. For~$F$ that are monotonic, Eq.~(\ref{eq:circle_map}) predicts an aperiodic increase of~$X_f$ similar to the dynamics in pulled waves~\citep{bak:devils_staircase}. However, once nonlinearities are strong enough to make~$F$ non-monotonic, the circle maps can lock into periodic oscillations. These oscillations arise from the fact that~$F_q$, which is the result of~$q$ consecutive applications of~$F$, develops an attractive fixed point~$F_q(X_f)=X_f$.\footnote{Non-monotonicity of~$F$ is required for an attracting fixed point in the~$F_q$ because any iterate of a monotonic~$F$ is monotonic.} The existence of such fixed points explains the periodicity, stability, and insensitivity to parameter changes of locked waves. The theory of mode locking in circle maps also predicts that not all parameter values that control the shape of~$F$ result in stable fixed point of~$F_q$~\citep{bak:devils_staircase, fat_fractal:prl, arnold:scaling}. In the context of range expansions, this means that velocity plateaus account for only part of the~$v(m)$ plot---the Devil's staircase is not complete. Therefore, at some values of~$m$ front propagation occurs without strict periodicity and velocity locking even though the expansions are not pulled. We termed such unlocked expansions pushed waves. Consistent with the above arguments, we found that pushed and locked invasions occur in DSDT models only in the presence of positive density-dependence in dispersal or growth. Negative density-dependence instead produces pulled invasions, which proceed in a smooth, continuous fashion in both discrete and continuous models.

Velocity locking could have important implications for the management of invasive species because the invariance of invasion rate under parameter variation could dramatically change how invaders respond to interventions. Indeed, a management strategy could be deemed ineffective when it does not result in any change of the invasion rate. Yet, a dramatic reduction in the rate of invasion may occur if the management effort is intensified beyond a point necessary to push the population to a new velocity plateau, say from~$v=1/2$ to~$v=1/4$ or even~$v=0$. 

Velocity locking also affects the dynamics and evolution of the invading population. We found that locked fronts propagate quasi-deterministically without exhibiting any diffusive wandering due to demographic or environmental fluctuations. In contrast, unlocked fronts have a nonzero effective diffusion constant that becomes especially large near velocity plateaus. While velocity locking does not suppress evolution of dispersal and growth, it nevertheless leads to important differences compared to the dynamics in continuous models. The analysis of reaction-diffusion equations suggests mutants are often selected on their expansion velocity~\citep{deforet:velocity_rule, korolev:arrest}. For locked invasion, mutations that do not change the expansion velocity can nevertheless be selected~(Fig.~\ref{fig:evolution}). Sequential accumulation of such mutation can shift the population to a new velocity plateau, which would appear as the case of both very rapid evolution and very strong epistasis.

Our results show that that velocity locking occurs in both one and two spatial dimensions and is extremely robust. It persists despite external perturbations, demographic noise, environmental fluctuations, and habitat heterogeneity. Velocity locking also does not require perfect discreteness of space and time; approximate periodicity in time and space is sufficient. Therefore, we expect that locked front should be easily observable in laboratory experiments and possibly in nature. The lack of observations to date might be attributed to the lack of awareness of mode locking phenomena among the scientists who study range expansions. Indeed, mode locking is not widely known outside nonlinear sciences, and the literature on population biology is dominated by continuous models, in which velocity locking cannot occur. 

While the main focus of our work has been on biological invasions, velocity locking could also have implications beyond ecology. Indeed, front propagation arises in quantum chromodynamics~\citep{marquet:qcd}, entanglement spreading~\citep{schachenmayer:entanglement, jurcevic:entanglement}, chemical kinetics~\citep{douglas:assembly_wave, pelce:curved_fronts}, biofilm growth~\citep{jacob:review}, cancer biology~\citep{gerlee:cancer_wave}, and population genetics of spatial~\citep{fisher:wave, kolmogorov:wave, lavrentovich:fixation} and well-mixed~\citep{tsimring:wave, rouzine:wave} populations. In many of these contexts, the spatio-temporal structure of the environment could be discrete or periodic and, therefore, support velocity locking together with all the unusual behaviors associated with locked fronts.

%*************************************************************************
% Methods
%*************************************************************************
\section{Methods}
\subsection{Simulations of CSCT, CSDT, and DSCT models}

\noindent For CSCT models, simulations in Fig.~\ref{fig:table_schematic}A were performed using the \texttt{pdepe} function in MatLab\textsuperscript{\textregistered} with error tolerance of~$10^{-3}$. Simulations in Fig.~\ref{fig:table_schematic}B were performed using finite-difference approach with discretization $\Delta x = 0.05$ and $\Delta t = 0.475 (\Delta x)^2$.
 
For DSCT models, we used the cubic~$g(c)$ defined in Eq.~(\ref{eq:cubic}) and solved~Eq.~(\ref{eq:dsct}) for~$2000$ patches using the standard Runge-Kutta method in MatLab\textsuperscript{\textregistered}. 

In all models, the habitat was empty initially, but the boundary conditions were chosen to mimic an expansion from a region where the population is well established. Specifically, population density was set to the carrying capacity on the left edge of the habitat. The boundary condition on the right edge was absorbing, but we did not run simulations until that boundary was reached. In all simulations, the velocity was obtained via least-square fitting of the front position to a linear function of time. 

\subsection{Simulations of DSDT models}
Simulations were performed as follows. Each generation, population densities were updated according to Eq.~(\ref{eq:dsdt}) with reflecting boundary conditions on both ends of the simulation box. The size of the simulation box contained at least~$100$ patches, and its position was periodically adjusted to ensure that the expansion front is centered. Prior to measuring the invasion velocities, we allowed the population to expand for at least~$10^4$ generations starting from a step-like initial condition. The velocities were then determined from a least-square linear regression of the position of the population front on time, during a period of at least another~$10^4$ generations. For stochastic simulations of DSDT models in Figs.~\ref{fig:fluctuations} and~\ref{fig:front_diffusion}, the front position was defined as the sum of the population densities in all patches normalized by the carrying capacity.

Because our main goal was to demonstrate that periodic front propagation arises due to velocity locking, we were careful to exclude other sources of oscillations. In particular, we considered~$f(c)$ without over-compensatory dynamics and limited the range of dispersal rate to~$[0,0.5]$. For higher values of~$m$, the migration matrix is no longer positive definite, which can trigger oscillations. As an example, consider the dynamics without growth for~$m=1$ with every even patch initially occupied and every odd patch initially empty.

We list parameters used in Figure~\ref{fig:table_schematic} here. In panel A for CSCT, $m=3$ and cubic growth with ~$K=1$, $g_0=1$, and $c_a=0.25$. For DSCT, $m=2$ and cubic growth with~$K=1$, $g_0=1$, and $c_a=-1.7$. For CSDT, $m=0.0204$ and piecewise-linear growth with~$K=1$, $r=0.2$, and $c^*=0.3$. For DSDT,~$m=0.25$ and piecewise-linear growth with $r=0.7$, $c^*=0.15$, $K=1$. In panel B For CSCT, cubic growth with ~$g_0=4$, $c_a=0.4$, and~$K=1$ for the strong Allee effect and~$g_0=0.25$, $c_a=-1.2$, and~$K=1$ for no Allee effect. For DSCT, cubic growth ~$g_0=1.1$, $c_a=0.25$, and~$K=1$ for the strong Allee effect and~$g_0=1.1$, $c_a=-1.1$, and~$K=1$ for no Allee effect. For CSDT, piecewise-linear growth with ~$r=0.5$, $c^*=0.3$, and~$K=1$ for the strong Allee effect and~$r=10/3$, $c^*=0.3$, and~$K=1$ for no Allee effect. For DSDT, piecewise-linear growth with~$r=0.7$, $c^*=0.25$, and~$K=1$ for strong Allee effect and ~$r=5.0$, $c^*=0.25$, and~$K=1$ for no Allee effect.

\subsection{Simulations of CSCT models with spatial and temporal periodicity in Fig.~\ref{fig:continuous}}
To handle the time and space dependent coefficients in Eq.~\eqref{eq:continuous}, simulations were performed with the following discretization scheme
\begin{equation}
\begin{split}
c_{t+\Delta t,x} =& c_{t,x} + \Delta t \bigg[ g_{t,x} \: c_{t,x}(1-c_{t,x} )(c_{t,x}-c^*) -j_{t,x} \: c_{t,x}  \\ & +  \left( D_{t,x+\frac{\Delta x}{2}}\: c_{t,x+\Delta x} \: + D_{t,x-\frac{\Delta x}{2}} \: c_{t,x-\Delta x}	\: -c_{t,x} \left( D_{t,x-\frac{\Delta x}{2}} + D_{t,x+\frac{\Delta x}{2}} \right) \right) \frac{1}{\left(\Delta x \right)^2}   	 \bigg]
\end{split}
\label{eq:continuous_discretization}
\end{equation}
where~$\Delta t$ and~$\Delta x$ were the time and space discretizations used for the numerical solution.

Runs were performed at two separate discretization values to ensure that results were independent of the choice of discretization. 

\appendix
\section{Velocities of pulled expansions}

\subsection{Velocity in pulled DSDT models}
Here, we derive the velocity of pulled expansions for DSDT models shown in Fig.~\ref{fig:pulled_works}. Our calculation closely follows that of van Saarloos~\citep{saarloos:review} and is given here for completeness.

When expansions are pulled, their properties are determined by the dynamics at the expansion front, where~$c_{t,x}$ is small and~$f(c)$ can be approximated as~$\rho c$ where~$\rho=\lim_{c\to0}f(c)/c$. In the absence of an Allee effect, this approximation yields the maximal possible growth rate; therefore, the bulk dynamics cannot push the population to expand at a higher rate than predicted by the linear approximation. For a weak Allee effect, the magnitude of the growth rate increase in the bulk determines whether the expansion is driven by its edge~(pulled waves) or its bulk~(pushed waves). For a strong Allee effect, the linear expansion results in~$\rho<0$, and the linear analysis is clearly unable to describe the expansion dynamics.

The linearization of Eq.~(\ref{eq:dsdt}) for DSDT models yields 

\begin{equation}
	c_{t+1,x} = \rho \left[ \frac{m}{2}c_{t,x-1} + (1-m)c_{t,x} + \frac{m}{2}c_{t,x+1} \right],
	\label{eq:dsdt_linear}
\end{equation}
\noindent with~$\rho=\lim_{c\to0}f(c)/c$.

This equation is solved through Fourier transforms defined as

\begin{equation}
	c_t(k) = \sum_x e^{-ikx}c_{t,x},
\end{equation}

\noindent with the following result

\begin{equation}
	c_t(k) = c_0(k) e^{t\ln\left(\rho[1-m(1-\cos k)]\right)},
\end{equation}

\noindent where~$c_0(k)$ is the Fourier transform of the initial population density.

We then shift to the comoving reference frame with a spatial coordinate~$z=x-vt$ and perform the inverse Fourier transform:

\begin{equation}
	c(t, z) = \int\frac{dk}{2\pi} c_0(k) e^{ikz} e^{t\left[ ikv + \ln(\rho[1-m(1-\cos k)]) \right]}.
	\label{eq:fourier_solution}
\end{equation}

\noindent This integral can be evaluated using the saddle point approximation~\citep{bender:orszag}, which is appropriate for large~$t$ when the transient dynamics are over. The saddle point approximation states that the integral is dominated by a region near~$k^*$, a particular value of~$k$ in the complex plane for which the complex derivative with respect to~$k$ of the power in the second exponent in Eq.~(\ref{eq:fourier_solution}) is zero:

\begin{equation}
	iv - \frac{m \sin k^*}{1 - m(1-\cos k^*)} = 0.
	\label{eq:saddle_condition}
\end{equation}

\noindent Since this is a complex equation, both real and imaginary parts need to be zero, which yields two real equations for the three unknowns: the real and imaginary parts of~$k^*$ and~$v$. The third required equation comes from the requirement that~$c(t,z)$ does not increase to infinity or diminish to zero at long times, which is due to the definition of a comoving reference frame. We then obtain the following condition:

\begin{equation}
	\Re\left\{ivk^* + \ln(\rho[1 - m(1-\cos k^*)])\right\} = 0.
	\label{eq:no_growth_condition}
\end{equation}

It is easy to see that Eqs.~(\ref{eq:saddle_condition}) and~(\ref{eq:no_growth_condition}) are satisfied only by purely imaginary~$k^*=i\kappa$, which yields the expected exponential decay of~$c(z)$ with~$z$ as~$e^{-\kappa z}$ from the first exponent in Eq.~(\ref{eq:fourier_solution}). The velocity~$v$ and asymptotic decay rate~$\kappa$ are then given by the following system of equations.

\begin{equation}
	v = \frac{\ln\left(\rho[1 + m (\cosh\kappa-1)]\right)}{\kappa},
	\label{eq:v}
\end{equation}

\begin{equation}
	v = \frac{m\sinh \kappa}{1 + m (\cosh\kappa-1)}.
	\label{eq:kappa}
\end{equation}

The second equation is equivalent to the requirement that~$\kappa$ minimizes~$v$ in the first equation. Thus, we can alternatively express~$v$ as

\begin{equation}
	v = \min_{\kappa>0}\left\{\frac{\ln\left(\rho[1 + m (\cosh\kappa-1)]\right)}{\kappa}\right\}.
	\label{eq:v_minAPP}
\end{equation}

\subsection{Velocity in pulled DSCT models}

The derivation for the pulled velocity of DSCT models mirrors the derivation above. The linearization of Eq.~(\ref{eq:dsct}) for DSCT models yields 

\begin{equation}
	\label{eq:dsct_linear}
	\frac{d c_x}{d t}=\frac{m}{2}(c_{x+1}-2c_x+c_{x-1})+ \rho c_x,
\end{equation} 
\noindent where~$\rho=\lim_{c\to0}g(c)/c$.

As above, we solve this equation via Fourier transform, inverse Fourier transform and moving to the comoving reference frame with~$z=x-vt$. This gives the analog of Eq.~(\ref{eq:fourier_solution})
\begin{equation}
	c_z(t) = \int\frac{dk}{2\pi} c_k(0) e^{ikz} e^{t\left[ ikv + \rho - m(1-\cos k) \right]}.
	\label{eq:fourier_solution_DSCT}
\end{equation}
Like above, we evaluate the integral via the saddle point approximation and use the requirement that~$c_z(t)$ does not increase to infinity or diminish to zero at long times. Thus, we arrive at the pulled velocity for DSCT models
\begin{equation}
	v = \min_{\kappa>0}\left\{\frac{\left( \rho + m (\cosh\kappa-1)\right)}{\kappa}\right\}.
	\label{eq:v_DSCT}
\end{equation}

\subsection{Velocity in pulled CSDT models}
The pulled velocity for CSDT models has been derived~\citep{kot:csdt}. For completeness, we reproduce their results using the same method as in the previous subsections. The linearization of Eq.~(\ref{eq:ide}) for CSDT models yields:

\begin{equation}
	\label{eq:ide}
	c_{t+1}(x)= r \int_{-\infty}^\infty Q(x-x')  c_t(x')dx',
\end{equation}
where~$r=\lim_{c\to0}f(c)/c$. 

We solve this equation via Fourier transform in space and then perform the inverse Fourier transform and a shift into the comoving reference frame with~$z=x-vt$:

\begin{equation}
	c_t(z) = \int\frac{dk}{2\pi} c_0(k) e^{ikz} e^{t\left[ ikv + \ln\left(\rho Q(k) \right)  \right]},
	\label{eq:fourier_solution_DSCT}
\end{equation}

where $Q(k)$ is the Fourier transform of the dispersal kernel. We then evaluate the integral via the saddle point approximation, using the requirement that~$c_z(t)$ does not increase to infinity or diminish to zero at long times. The result reads

\begin{equation}
	v = \min_{\kappa>0}\left\{  \frac{  \ln\left[ \rho \, Q(-i \kappa) \right] }{\kappa}\right\}
	\label{eq:v_DSCT}
\end{equation}

Note that the Fourier transform of the dispersal kernel evaluated at $-i \kappa$ is equivalent to calculating the moment generating function of~$Q(x)$, which was used by Ref.~\citep{kot:csdt}.

For the piecewise-linear model with Laplace kernel used in simulations, the above result reduces to
\begin{eqnarray}
v =  \min_{\kappa>0}\left\{  \frac{  \ln\left[ \frac{\rho}{1-m \kappa^2} \right] }{\kappa}\right\}
\end{eqnarray}

The pulled velocities for all four classes of models are compared to simulations in Fig.~\ref{fig:table_schematic}B.

\section{Expansions locked at~$v=\frac{1}{2}$ in the piecewise-linear model}
We now show that~$v(m)$ indeed exhibits exact plateaus by explicitly finding traveling-wave solutions of Eq.~(\ref{eq:dsdt}) with~$v=1/2$ for the piecewise-linear model. Since velocity locking occurs only in the presence of an Allee effect, we will assume that~$rc^*<K$. Our strategy is analogous to that deployed for partial differential equations with piecewise-linear growth~\citep{korolev:wave_splitting}: We solve in each region of~$c$ where~$f(c)$ is linear and then match the solutions. For~$c>c^*$, the solution is~$c_{t,x}=K$, i.e. the bulk of the wave is at the carrying capacity. For~$c<c^*$, the dynamics are described by Eq.~(\ref{eq:dsdt_linear}), which admits exponential solutions of the form

\begin{equation}
	c_{t,x} = C e^{-\lambda(x-vt)},
	\label{eq:exponential_solution}
\end{equation}

\noindent where~$C$ is the amplitude of the solution that should be determined from matching to the solution behind the front, and~$\lambda$ is the spatial decay rate chosen to satisfy Eq.~(\ref{eq:dsdt_linear}) with~$v=1/2$, i.e. we require that

\begin{equation}
	e^{\lambda/2} = r[1+m(\cosh\lambda-1)].
	\label{eq:lambda}
\end{equation}

To analyze the nature of solutions to this equation, it is convenient to define~$l=e^{\lambda/2}$ and rewrite the equation as follows:
\begin{equation}
	m = \frac{2}{r}\frac{l^2 (l-r)}{(l^2-1)^2}.
\label{eq:lambda_existence}
\end{equation}

For~$r\le1$, i.e. for the strong Allee effect, the right hand side is monotonically decaying from~$+\infty$ to~$0$ as~$l$ increases from its minimal value of~$1$ to~$+\infty$. As a result, there is a unique solution~$\lambda(m)$ of Eq.~(\ref{eq:lambda}). When the Allee effect is weak~($r>1$), the right hand side of equation Eq.~(\ref{eq:lambda_existence}) increases from~$-\infty$ to a maximum~($m_e$) and then declines to~$0$ as~$l$ increases from~$0$ to~$+\infty$. Hence, there may be two solutions when the migration rate is below that maximum~($m<m_e$) or no solutions when it is above the maximum~($m>m_e$). The value of~$m_e$ is defined as
\begin{equation}
	m_e = \max_{l\in(1,+\infty)}\left\{ \frac{2}{r}\frac{l^2 (l-r)}{(l^2-1)^2} \right\}.
\label{eq:m_e}
\end{equation}

Clearly, for~$r>1$, Eq.~(\ref{eq:m_e}) sets an upper bound on the migration rates consistent with~$v=1/2$. We also note that, when Eq.~(\ref{eq:lambda}) has two solutions, only the larger one corresponds to a pushed expansion. This situation is completely analogous to that for reaction-diffusion equations discussed in Ref.~\citep{saarloos:review}. The smaller~$\lambda$ corresponds to solutions arising from initial conditions that decay very slowly at~$+\infty$ and therefore lack biological realism since all expansions are started by a population completely confined to a finite region of space.

With~$\lambda$ defined by the largest solution of Eq.~(\ref{eq:lambda}), the solution for~$c_{t,x}$ is then given by the following ansatz:

\begin{equation}
	c_{t,x} = \left\{
	\begin{aligned}
		& K, \;\; x-t/2 \le x_0 \\
		& Ke^{-\lambda(x-t/2)}, \;\; x-t/2 > x_0,
	\end{aligned}
	\right.
	\label{eq:ansatz}
\end{equation}

\noindent where~$x_0$ determines the initial position of the front. In the following, we assume that~$x_0=0$ for simplicity. Note that we set~$C=K$ due to the matching requirement. Indeed, for the exponential profile to be preserved in the nonlinear model, the right most point of the bulk solution must also satisfy Eq.~(\ref{eq:lambda}) in the generation following the update that transforms the left most point of the front solution to the right most point of the bulk solution. In other words,~$C=K$ ensures that the density of the left most point below the carrying capacity is always consistent with the exponential decay at the front.

Three other conditions are necessary for the ansatz to hold. (i) The right most point of the bulk region should not fall below~$c^*$ following migration; otherwise, it will fall below the carrying capacity and would not be able to send enough migrants to its neighbor that needs to reach the carrying capacity in the next generation. (ii) The density of the first point below the carrying capacity must not increase above~$c^*$ after migration following the generations when~$x-t/2$ is an integer; otherwise the bulk region will advance by one step every generation instead of every other generation. (iii) The opposite must be true following the generations when~$x-t/2$ is a half integer to ensure that the bulk region advances by one patch every two generations. Below we state these conditions as appropriate inequalities and identify the region of~$m\in(m_{\mathrm{min}}, m_{\mathrm{max}})$ that is consistent with~$v=1/2$, i.e. we compute the location of the corresponding velocity plateau.

First condition:

\begin{equation}
	m <  \frac{2 (1 - c^*/K)}{1-e^{-\lambda}}.
	\label{eq:condition_1}
\end{equation}

Second condition:

\begin{equation}
	e^{-\lambda/2}<rc^*/K.
	\label{eq:condition_2}
\end{equation}

Third condition:

\begin{equation}
	e^{-\lambda/2} + \frac{m}{2} \left( 1 + e^{-3\lambda/2} - 2e^{-\lambda/2} \right) > c^*/K.
	\label{eq:condition_3}
\end{equation}

In our simulations, the first condition was never violated, so it might be unnecessary at least for~$m<0.5$. The last two conditions can be used to obtain analytical expressions for~$m_{\mathrm{min}}$ and~$m_{\mathrm{max}}$ by changing inequalities to equalities, which are satisfied at these ``critical'' values of~$m$, using Eq.~(\ref{eq:lambda}), performing a change of variables~$l=e^{\lambda/2}$, and solving for the critical values of~$l$. Some of these expressions, however, are sufficiently complex and hard to gain intuition from; moreover, one still needs to ensure that the critical values of~$l$ correspond to the largest solution of Eq.~(\ref{eq:lambda}). In consequence, it is often much more straightforward to find~$\lambda(m)$ numerically from Eq.~(\ref{eq:lambda}) and verify that all three conditions are satisfied.

For completeness, we also provide the solutions for the critical values of~$m$ discussed above. From Eqs.~(\ref{eq:lambda}) and (\ref{eq:condition_2}), we conclude that

\begin{equation}
	m < 2\frac{c^*}{K} \frac{1- r^2 c^*/K}{(1-(rc^*/K)^2)^2},
	\label{eq:m_max}
\end{equation}

\noindent provided~$\lambda = -2\ln(rc^*/K)$ is the largest root of Eq.~(\ref{eq:lambda}). From Eqs.~(\ref{eq:lambda}) and (\ref{eq:condition_3}), we conclude that

\begin{equation}
	m > \frac{2}{l_c - 1}\frac{1-rc^*/K}{r},
	\label{eq:m_min}
\end{equation}

provided~$\lambda = 2\ln(l_c)$ is the largest root of Eq.~(\ref{eq:lambda}), and~$l_c$ is the positive root of the following equation:

\begin{equation}
	l^3\left(\frac{r c^*/K}{1-rc^*/K}\right) - l^2\left(1+ \frac{r}{1-rc^*/K}\right) + l + 1 = 0.
	\label{eq:l_cubic}
\end{equation}

\noindent When~$\lambda = 2\ln(l_c)$ is the only or the smallest root of Eq.~(\ref{eq:lambda}), there is no velocity locking. Further, note that in addition to Eq.~(\ref{eq:m_max}), $m$ is bounded above by~$m_{e}$ when~$r>1$ and the requirement that~$m\le0.5$ in our simulations.

\section*{Acknowledgements}
Preliminary work was carried out by Vipul Vachharajani, Eugene Yurtsev, and Jeff Gore, who observed velocity locking in simulations. This work was partially supported by a grant from the Simons Foundation (\#409704, Kirill S. Korolev), by the Cottrell Scholar Award (\#24010, Kirill S. Korolev), and by a grant from Moore foundation~(\#6790.08, Kirill S. Korolev). Simulations were carried out on Shared Computing Cluster at Boston University.

%*************************************************************************
% Bibliography
%*************************************************************************
\bibliography{velocity_locking_final}
\bibliographystyle{vancouver}

\end{document}